\documentclass[pra,aps,twocolumn,superscriptaddress,nofootinbib,nopacs]{revtex4} 

\usepackage{amsmath}
\usepackage{amssymb}
\usepackage{amsfonts}
\usepackage{bm}
\usepackage{braket}
\usepackage{color}
\usepackage{dcolumn}
\usepackage{epsfig}
\usepackage{gensymb}
\usepackage{graphicx}
\usepackage{lmodern}
\usepackage{mathrsfs}
\usepackage{mathtools}
\usepackage{pst-all}
\usepackage{psfrag}
\usepackage{units}
\usepackage[colorlinks,linkcolor=blue,citecolor=blue,urlcolor=blue,hyperindex,driverfallback=dvipdfm]{hyperref}

\def\ii{{\rm i}}  \def\ee{{\rm e}}
\def\Ree{{\rm Re}}  
\def\rb{{\bf r}}      
\def\xx{\hat{\bf x}}  \def\yy{\hat{\bf y}}  \def\zz{\hat{\bf z}}  \def\nn{\hat{\bf n}}
\def\nn{\hat{\bf n}}  \def\eh{\hat{\bf e}}
  
\def\kb{{\bf k}}  \def\kpar{k_\parallel}  \def\kparb{{\bf k}_\parallel}
    
\def\Eb{{\bf E}}    \def\Hb{{\bf H}}  
\def\pb{{\bf p}}  
  
    \def\EF{{E_{\rm F}}}
\def\lamp{\lambda_{\rm p}}      
\def\wp{\omega_{\rm p}}
  
  \def\mb{{\bf m}}  \def\bb{{\bf b}}
  
\def\aE{\alpha_{\rm E}}  \def\aM{\alpha_{\rm M}}
\def\trp{\tilde{r}_{\kpar{\rm p}}}  \def\trs{\tilde{r}_{\kpar{\rm s}}}  \def\epsilonh{\epsilon_{\rm h}}

\begin{document} 
\title{Complete Coupling of Focused Light to Surface Polaritons}


\author{Eduardo~J.~C.~Dias}
\affiliation{ICFO-Institut de Ciencies Fotoniques, The Barcelona Institute of Science and Technology, 08860 Castelldefels (Barcelona), Spain}
\author{F.~Javier~Garc\'{\i}a~de~Abajo}
\email{javier.garciadeabajo@nanophotonics.es}
\affiliation{ICFO-Institut de Ciencies Fotoniques, The Barcelona Institute of Science and Technology, 08860 Castelldefels (Barcelona), Spain}
\affiliation{ICREA-Instituci\'o Catalana de Recerca i Estudis Avan\c{c}ats, Passeig Llu\'{\i}s Companys 23, 08010 Barcelona, Spain}


\begin{abstract}
Surface polaritons display short wavelengths compared to propagating light, thus enabling large spatial concentration and enhancement of electromagnetic energy. However, this wavelength mismatch is generally accompanied by poor light-to-polaritons coupling that limits potential applications in areas such as optical sensing and optoelectronics. Here, we address this problem by demonstrating that a small scatterer placed at a suitable distance from a planar surface can produce complete coupling of a focused light beam to surface polaritons. We present rigorous theoretical results for light coupling to plasmons in silver films and graphene, as well as phonon polaritons in hexagonal boron nitride films. We further formulate detailed general prescriptions on the beam profile and particle response that are required to achieve maximum coupling, which we supplement by analytical calculations for dipolar scatterers and finite-size particles. Our results open a practical route to circumvent the long-standing photon-polariton wavelength mismatch problem in nanophotonics.
\end{abstract}

\maketitle
\date{\today} 


\section{Introduction}

Light manipulation at deep subwavelength scales has attracted considerable interest in recent years because of its applications in diverse areas of nanophotonics, ranging from optical sensing \cite{LNL1983,AHL08,JHE08,KEP09,paper125,ZBH14} to light harvesting \cite{CP08,AP10}, photodetection \cite{LLK13,KMA14,XML09,paper315}, and nonlinear optics \cite{VDC01,SJ04,DSK08,PN08,GTG10,KZ12,paper337}. Surface polaritons are pivotal in this effort, as they can exhibit in-plane wavelengths that are substantially smaller than the wavelength of light propagating at the same frequency \cite{paper283,WLG15,LGA17,paper335}. This concept has been pushed to the atomic-scale limit with the emergence of two-dimensional materials capable of sustaining different types of surface polaritons \cite{paper283}, including graphene plasmons \cite{paper176,WLG15,LGA17}, atomically-thin metal plasmons \cite{MRH99,RNP08,paper335,paper326}, and optical phonons in hexagonal boron nitride (hBN) \cite{DFM14}, which have proved to be promising for applications in optical sensing \cite{paper256,paper319}, infrared spectrometry \cite{YLL12,paper315}, and nonlinear nanophotonics \cite{CHC16,paper337}.

Achieving good coupling between light and polaritons is critically important for practical applications involving external light sources. Advances in this direction have relied on the use of optical cavities, such as nanoscale tips, which have become a popular choice to achieve large near-field enhancements and increased coupling to surface modes for imaging and spectroscopy purposes \cite{HTK02,HOH08,WDR14}. However, this approach still lies far from the ideal situation of complete light-to-polariton coupling at designated spatial locations. In this context, it has recently been shown that coupling through a near-the-surface dipolar scatterer results in a light-to-polaritons scattering cross section $\sim\lamp^3/\lambda_0$, expressed in terms of the polariton and light wavelengths, $\lamp$ and $\lambda_0$, respectively \cite{paper331}. The coupling cross section is consequently small compared with the area of a diffraction-limited light spot $\sim\lambda_0^2$, thus resulting in a poor photon-to-polariton conversion ratio $\sim\lamp^3/\lambda_0^3$.

When a lossless dipolar scatterer is placed in a homogeneous dielectric environment, it can produce complete extinction of a focused light beam \cite{ZMS08}. A practical realization of this idea led to the demonstration of substantial extinction by a single two-level molecule, which was required to be held under cryogenic conditions to prevent inelastic losses \cite{RWL12}, followed by a recent demonstration of coupling between a molecule and a plasmonic nanoparticle \cite{ZGR20}. Now, by bringing the scatterer close to a surface in order to assist coupling to surface modes, it is precisely this coupling that unavoidably introduces losses from the point of view of the scatterer, an effect that leads in turn to a reduction in the scattering strength and the associated coupling efficiency \cite{paper331}. We now ask the question whether the efficiency of light-to-polaritons coupling assisted by a small scatterer can be increased by separating it from the surface, so that it is positioned at an optimum finite distance defined by the compromise between an efficient interaction with the evanescent fields associated with surface modes and the resulting reduction in surface-related inelastic losses that quench the strength of the scatterer. This idea resembles the so-called critical coupling conditions, whereby complete absorption of a light plane wave by a structured planar surface is produced when radiative and nonradiative loss rates are made equal \cite{HM1976,paper107,paper182}. We expect that the scatterer-surface distance provides a knob to vary the balance between radiative and nonradiative loss processes, but the overall efficiency of the proposed scheme remains uncertain.

In this work, we demonstrate that a small scatterer placed at an optimum distance in front of a planar surface can produce complete coupling from a suitably shaped focused light beam into surface modes. We based our results on rigorous analytical theory for small dipolar scatterers, and further corroborate their validity through numerical simulations for finite-size silicon particles, which act as nearly-perfect scatterers at their dipolar Mie resonances and can produce $>90\%$ absorption of a focused light beam by a silver surface. We offer detailed general prescriptions on the required angular profile of the incident light beam and the characteristics of the scatterer, which depend on the composition and optical properties of the surface material. We illustrate this concept with examples of optimum coupling to planar dielectric waveguide modes, as well as phonons in hBN and plasmons in graphene and ultrathin metallic films. This study opens a new route toward complete coupling between light and surface polaritons based on realistic structures incorporating an engineered structure placed in front of the surface. Additionally, our results can be readily applied to two-level quantum emitters, such as optically-trapped atoms and molecules in the vicinity of a polariton-supporting surface, which can equally assist both complete optical coupling and strong polariton-mediated interaction among emitters placed at designated positions.

\begin{figure*}[ht]
\centering
\includegraphics[width=0.8\textwidth]{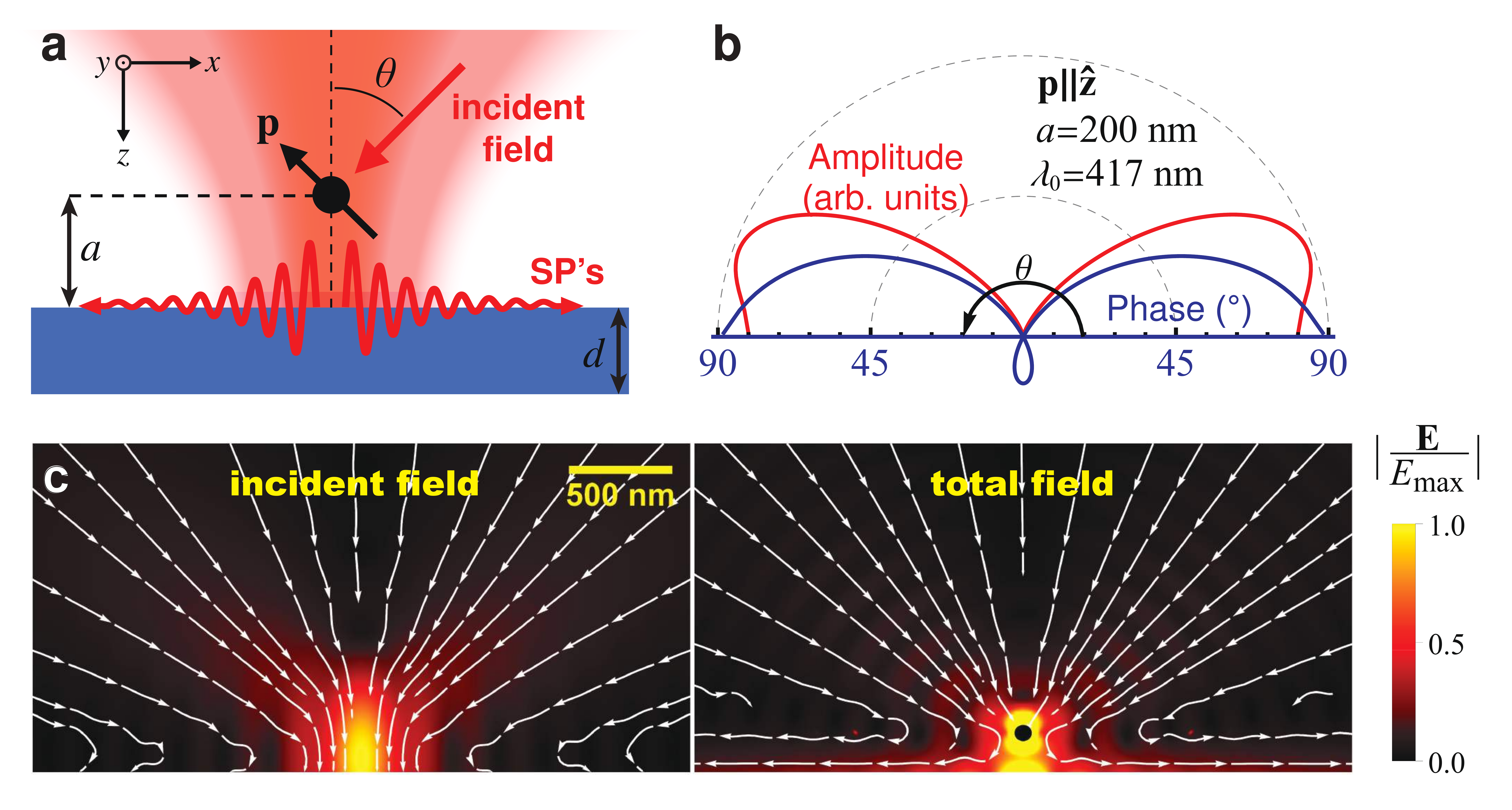}
\caption{Optimum light coupling to surface polaritons mediated by a small particle. (a) Illustration of the configuration under consideration. A small particle, described through its induced dipole $\pb$, is placed at a distance $a$ from the planar surface of a homogeneous film. A light beam with optimized angular profile couples maximally to polaritons supported by the film. (b) Angular phase and amplitude  profile of the optimum light beam needed to achieve complete coupling to a semi-infinite silver surface with axially-symmetric p-polarized light, which renders $\pb\parallel\zz$. We take a particle-surface separation $a=200\,$nm and a light wavelength $\lambda_0=417\,$nm. (c) Electric field amplitude (color plots) and Poynting vector field lines (white arrows) of the incident focused beam in the absence (left) and presence (right) of the surface+scatterer system.}
\label{Fig1}
\end{figure*}

The type of system under consideration is sketched in Fig.\ \ref{Fig1}(a), consisting of a scatterer situated at a distance $a$ from a homogeneous planar surface. Figure\ \ref{Fig1}(b) shows an example of the angular profile of the incident beam that is required to achieve complete absorption by a semi-infinite silver surface (see below for details of the calculation), whereas Fig.\ \ref{Fig1}(c) portrays the incident (left) and total (right) electric near-field amplitude. We superimpose on the latter the associated Poynting vector field lines (white arrows), which are clearly redirected along the surface in the presence of the particle, indicating that absorption of the incident light is dominated by coupling to surface plasmons.

\section{Extremal scattering properties of a small particle}
\label{sec2}

Before introducing a plane surface, we consider a particle in a homogeneous environment and concentrate on externally supplied monochromatic light of frequency $\omega$, writing the optical electric field at a position $\rb$ and time $t$ as $\Eb(\rb,t)=2\Ree\left\{\Eb(\rb)\ee^{-\ii\omega t}\right\}$. The field amplitude in a constant-$z$ plane fully contained inside a homogeneous medium of permittivity $\epsilonh$ can be rigorously decomposed into plane waves as
\begin{align}
\Eb(\rb)=\sum_{\nu=\pm}\sum_{\sigma={\rm s,p}}\int\frac{d^2\kparb}{(2\pi)^2}
\,\beta^\nu_{\kparb\sigma}
\,\eh^\nu_{\kparb\sigma}\,\ee^{\ii\kb'^\nu\cdot\rb},
\label{Efield}
\end{align}
where the integral extends over parallel wave vectors $\kparb=(k_x,k_y)$ with $\kpar<k'=k\sqrt{\epsilonh}$ and $k=\omega/c$; the sums run over polarizations $\sigma\!=$s,\,p (or equivalently, TE and TM, respectively) and propagation toward positive ($\nu=+$) and negative ($\nu=-$) $z$ directions; $\kb'^\pm=\kparb\pm k'_z\zz$ are the respective light wave vectors of $z$ component $k'_z=\sqrt{k'^2-\kpar^2+\ii0^+}$; $\eh^\pm_{\kparb{\rm s}}=(1/\kpar)(-k_y\xx+k_x\yy)$ and $\eh^\pm_{\kparb{\rm p}}=(1/k'\kpar)(\pm k'_z\kparb-\kpar^2\zz)$ are the corresponding unit polarization vectors; and $\beta^\nu_{\kparb\sigma}$ are  expansion coefficients. Upon direct integration of the Poynting vector, we find the power transported by the field across the $z$ plane under consideration to be $\mathcal{P}^+-\mathcal{P}^-$ with
\begin{align}
\mathcal{P}^\pm=\frac{c\sqrt{\epsilonh}}{(2\pi)^3}\sum_{\sigma={\rm s,p}}\int d^2\kparb
\,\frac{k'_z}{k'}\,\left|\beta^\pm_{\kparb\sigma}\right|^2
\label{Power}
\end{align}
representing the power flowing upwards ($+$) and backwards ($-$) with respect to the $z$ axis.

\subsection{Maximum focal field, scattering, and absorption by a point particle in a homogeneous medium}
\label{maximalhomo}

We first explore the light profile needed to maximize the intensity of the electric field component along a direction determined by the complex unit vector $\nn$ (normalized as $\nn\cdot\nn^*=1$) at a focal point $\rb=0$ within the $z=0$ plane for fixed total incident power $\mathcal{P}^{\rm inc}=\mathcal{P}^++\mathcal{P}^-$. By imposing the vanishing of the functional derivatives of $\delta\big\{|\nn^*\cdot\Eb(0)|^2/\mathcal{P}^{\rm inc}\big\}/\delta\beta^{\nu*}_{\kparb\sigma}=0$, we find $\nn\cdot\eh_{\kparb\sigma}^{\nu}\;\mathcal{P}^{\rm inc}-\nn\cdot\Eb^*(0)\,(c/2\pi k)\,k'_z\beta_{\kparb\sigma}^\nu=0$, which leads to the optimum solution
\begin{align}
\beta^\nu_{\kparb\sigma}\propto\frac{1}{k'_z}\nn\cdot\eh^{\nu*}_{\kparb\sigma} \;\theta(k'-\kpar),
\label{betaE}
\end{align}
where the step function limits the coefficients to propagating waves with $\kpar<k'$ inside the light cone of the homogeneous dielectric medium. Now, inserting Eq.\ (\ref{betaE}) into Eqs.\ (\ref{Efield}) and (\ref{Power}), we find the maximum ratio
\begin{align}
{\rm max}\left\{\frac{|\Eb(0)|^2}{\mathcal{P}^{\rm inc}}\right\}=\frac{4\sqrt{\epsilonh}\,\omega^2}{3c^3}.
\label{maxE}
\end{align}
We remark that the focal field obtained by applying this prescription is fully parallel to $\nn$ [this follows from applying the identity $\sum_\sigma\eh^\nu_{\kparb\sigma}\otimes\eh^\nu_{\kparb\sigma}=\mathcal{I}-\kb'^\nu\otimes\kb'^\nu/k'^2$ to the evaluation of Eq.\ (\ref{Efield})], and furthermore, the ratio in Eq.\ (\ref{maxE}) is independent of the chosen orientation of $\nn$. For a beam propagating along positive $z$'s only (i.e., imposing $\beta^-_{\kparb\sigma}=0$), we still find the solution (\ref{betaE}) for the optimum field coefficients, but the ratio is halved to $2\sqrt{\epsilonh}\,\omega^2/3c^3$, although it still exceeds by a factor of $\approx1.43$ the value obtained for a diffraction-limited Gaussian beam \cite{paper331}, corresponding to the choice $\beta^+_{\kparb\sigma}\propto\nn\cdot\eh^+_{\kparb\sigma}\sqrt{k'_z/k'}$ \cite{NH06}.

Analogous results are obtained for the maximum focal magnetic field $|\nn^*\cdot\Hb(0)|$. Combining Eq.\ (\ref{Efield}) with Amp\`ere's law, $\Hb=\nabla\times\Eb/\ii k$, we find
\begin{align}
\Hb(\rb)=\sqrt{\epsilonh}\sum_{\nu=\pm}\sum_{\sigma={\rm s,p}}\int\frac{d^2\kparb}{(2\pi)^2}
\,\xi_\sigma\,\beta^\nu_{\kparb\sigma}
\,\eh^\nu_{\kparb\tilde\sigma}\,\ee^{\ii\kb'^\nu\cdot\rb},
\label{Hfield}
\end{align}
where $\tilde\sigma=$p,s when $\sigma\!=$s,p, while $\xi_{\rm p}=1$ and $\xi_{\rm s}=-1$. Following a similar procedure as above, we find the optimum field coefficients
\begin{align}
\beta^\nu_{\kparb\sigma}\propto\frac{\xi_\sigma}{k'_z}\nn\cdot\eh^{\nu*}_{\kparb\tilde\sigma} \;\theta(k'-\kpar)
\label{betaM}
\end{align}
needed to maximize $|\nn^*\cdot\Hb(0)|$ for fixed incident power (notice that p and s polarizations now have opposite signs due to the $\xi_\sigma$ factor), which lead to a ratio of magnetic field intensity to incident power $4\epsilonh^{3/2}\,\omega^2/3c^3$ [i.e., a factor of $\epsilonh$ larger than the result of Eq.\ (\ref{maxE}) for electric field maximization].

Incidentally, following this approach we find that the ratio between the electromagnetic energy density at the focus $\mathcal{U}=\left[\epsilonh\,|\Eb(0)|^2+|\Hb(0)|^2\right]/8\pi$ and the incident power is automatically maximized for any linear combination of the external light coefficients given by Eqs.\ (\ref{betaE}) and (\ref{betaM}), leading to
\begin{align}
{\rm max}\left\{\frac{\mathcal{U}}{\mathcal{P}^{\rm inc}}\right\}=\epsilonh^{3/2}\,\omega^2/6\pi c^3,
\label{Umax}
\end{align}
which is half of a previously reported upper bound for arbitrary fields \cite{B1986}, in which the constrain imposed by interference between magnetic and electric components was ignored. Because the expansion of Eq.\ (\ref{Efield}) is fully general for arbitrary propagating fields, the result of Eq.\ (\ref{Umax}) constitutes an absolute upper bound of $\mathcal{U}/\mathcal{P}^{\rm inc}$.

\subsection{Limits to absorption and scattering}

We now consider a small particle placed at $\rb=0$, whose optical response is dominated by an electric polarizability $\aE$ along a direction $\nn$. In response to external illumination, a dipole
\begin{align}
\pb=\aE\left[\nn^*\cdot\Eb(0)\right]\,\nn
\label{pb}
\end{align}
is induced, which in turn produces a scattered electric field given by Eq.\ (\ref{Efield}) with $\beta^\pm_{\kparb\sigma}$ replaced by \cite{NH06}
\begin{align}
\beta^\pm_{\kparb\sigma,{\rm dip}}=\frac{2\pi\ii k^2}{k'_z}\,\pb\cdot\eh^\pm_{\kparb\sigma},
\label{betadip}
\end{align}
where $\kpar$ is no longer limited to the light cone.

Interestingly, these coefficients have the same form as Eq.\ (\ref{betaE}), so for a suitably chosen polarizability $\aE$, scattering by the particle can completely suppress transmission of the incident light, which is therefore fully reflected. This happens for instance for a beam propagating toward positive $z$'s with $\beta^+_{\kparb\sigma}=-\beta^+_{\kparb\sigma,{\rm dip}}$ and $\beta^-_{\kparb\sigma}=0$; indeed, by inserting these coefficients into Eq.\ (\ref{Efield}), and this in turn into Eq.\ (\ref{pb}), applying the identity $\sum_\sigma\int_{\kpar<k'}(d^2\kparb/k_z')(\nn\cdot\eh^+_{\kparb\sigma})(\nn^*\cdot\eh^+_{\kparb\sigma})=4\pi k'/3$, we find the condition $-\aE^{-1}=2\ii k'^3/3\epsilonh$ for perfect beam reflection. Now, we remind that the optical theorem imposes the constraint ${\rm Im}\{-\aE^{-1}\}\ge2k'^3/3\epsilonh$ \cite{V1981}, with the equality implying a lossless scatterer (i.e., when radiative losses account for the full value of ${\rm Im}\{-\aE^{-1}\}$); in addition to this, our scatterer must be resonant (i.e., ${\rm Re}\{-\aE^{-1}\}=0$). We conclude that a focused beam with the angular profile of a dipolar far field [Eq.\ (\ref{betaE})] is completely reflected by a lossless resonant dipolar scatterer, as previously predicted using arguments based on numerical simulations \cite{ZMS08}.

The condition for complete absorption (i.e., full cancellation of the incident field by the induced-dipole field) is given by $\beta^\pm_{\kparb\sigma}=-\beta^\pm_{\kparb\sigma,{\rm dip}}$, which upon insertion again into Eq\ (\ref{Efield}) and comparison with Eq.\ (\ref{pb}), leads to $-\aE^{-1}=4\ii k'^3/3\epsilonh$. Full absorption is then possible for a symmetric, dipole-like incident field (coming from both positive and negative $z$'s) interacting with a lossy resonant particle. The required value of ${\rm Im}\{-\aE^{-1}\}$ is then twice the minimum imposed by the optical theorem, therefore entailing an equal contribution of radiative and inelastic losses (the above mentioned critical coupling condition). If we remove the backward component of the incident field (i.e., setting $\beta^-_{\kparb\sigma}=0$) and repeat the above analysis, we find a maximum extinction (i.e., depletion of the incident beam due to both absorption and back scattering) of 50\%, provided the incident beam has a dipolar profile [i.e., $\beta^+_{\kparb\sigma}$ is shaped as in Eq.\ (\ref{betaE})] and the particle polarizability also satisfies $-\aE^{-1}=4\ii k'^3/3\epsilonh$.

This analysis can be repeated for a particle responding through a dipolar magnetic polarizability $\aM$, equally leading to complete absorption and reflection if $-\aM^{-1}=4\ii k'^3/3\epsilonh$ and $2\ii k'^3/3\epsilonh$, respectively, provided the incident light has the magnetic-dipole profile given by Eq.\ (\ref{betaM}). For completeness, we note that the scattered field then has components $\beta^\pm_{\kparb\sigma,{\rm dip}}=(2\pi\ii\xi_\sigma k^2/k'_z\sqrt{\epsilonh})\,\mb\cdot\eh^\pm_{\kparb\tilde\sigma}$, where $\mb$ is the induced magnetic dipole.

\section{Particle near a planar surface}

We consider a particle placed at $\rb=0$ near a planar surface defined by the $z=a>0$ plane and illuminated with only upward waves of coefficients $\beta^+_{\kparb\sigma}$ [i.e., $\beta^-_{\kparb\sigma}=0$, see Fig.\ \ref{Fig1}(a)]. The particle is also affected by field components reflected from the surface with downward coefficients $\beta^-_{\kparb\sigma}=\left(\beta^+_{\kparb\sigma}+\beta^+_{\kparb\sigma,{\rm dip}}\right)\,\tilde{r}_{\kpar\sigma}$, where $\beta^+_{\kparb\sigma,{\rm dip}}$ is defined in Eq.\ (\ref{betadip}),
\begin{align}
\tilde{r}_{\kpar\sigma}=r_{\kpar\sigma}\ee^{2\ii k'_za},
\nonumber
\end{align}
$r_{\kpar\sigma}$ are Fresnel's reflection coefficients, and the exponential accounts for propagation back and forth across the particle-surface distance $a$ in the embedding dielectric (i.e., $\tilde{r}_{\kpar\sigma}$ are the effective reflection coefficients referred to the plane of the particle, $z=0$). The field at the particle now becomes
\begin{align}
\Eb(0) = \sum_{\sigma} \int \frac{d^2 \kparb}{(2\pi)^2} \left\{ \beta_{\kparb\sigma}^+\,\bb_{\kparb\sigma}+ \tilde{r}_{\kpar\sigma}\,\beta^{+}_{\kparb\sigma,{\rm dip}}\,\eh_{\kparb\sigma}^{-} \right\},
\label{Edip0}
\end{align}
where
\begin{align}
\bb_{\kparb\sigma}=\eh^+_{\kparb\sigma}\!+\tilde{r}_{\kpar\sigma}\eh^-_{\kparb\sigma}.
\label{bb}
\end{align}
Inserting Eq.\ (\ref{betadip}) into Eq.\ (\ref{Edip0}), and this in turn into Eq.\ (\ref{pb}), we find the self-consistent particle dipole
\begin{align}
\pb=\frac{1}{1/\aE-\mathcal{G}}\cdot\sum_{\sigma={\rm s,p}}\int\frac{d^2\kparb}{(2\pi)^2}\,\beta^+_{\kparb\sigma}\,\bb_{\kparb\sigma},
\label{pbsurf}
\end{align}
where $\mathcal{G}=(\ii k^2/2\pi)\int (d^2\kparb/k'_z)
\sum_{\sigma}\tilde{r}_{\kpar\sigma}\,\eh^-_{\kparb\sigma}\otimes\eh^+_{\kparb\sigma}$ is a $3\times3$ Green tensor that accounts for the image dipole self-interaction. By using the explicit expressions given in Sec.\ \ref{sec2} for the polarization vectors, we find that $\mathcal{G}$ is diagonal and its elements $\mathcal{G}_{xx}=\mathcal{G}_{yy}\equiv\mathcal{G}_\parallel$ and $\mathcal{G}_{zz}\equiv\mathcal{G}_\perp$ reduce to
\begin{align}
\left[\begin{array}{c} \mathcal{G}_\parallel \\ \mathcal{G}_\perp \end{array} \right]
=\frac{\ii}{2\epsilonh}\int_0^\infty\frac{\kpar d\kpar}{k'_z}
\left[\begin{array}{c} k'^2\,\trs-k_z'^2\,\trp \\ 2\kpar^2\,\trp \end{array} \right].
\label{G}
\end{align}
In what follows we assume that the polarizability has the same symmetry as $\mathcal{G}$, with only $\alpha_{{\rm E}\parallel}$ and $\alpha_{{\rm E}\perp}$ nonzero components along directions $x$-$y$ and $z$, respectively, such that the induced particle dipole $\pb=p\nn$ is oriented either parallel or perpendicular to the planar surface.

\subsection{Maximum focal intensity in the presence of a planar surface}

The maximal focal field investigated in Sec.\ \ref{maximalhomo} is modified by the presence of materials (e.g., large field enhancement is produced by coupling to tightly confined plasmons \cite{BS03}). In the present context, a planar surface contributes to the field with reflected components, so for a beam propagating only along positive $z$ directions (i.e., $\beta^-_{\kparb\sigma}=0$), we can repeat the analysis of Sec.\ \ref{maximalhomo} with $\eh^+_{\kparb\sigma}$ replaced by $\bb_{\kparb\sigma}$, from which we find the condition
\begin{align}
\beta^+_{\kparb\sigma}\propto\frac{1}{k'_z}\nn\cdot\bb^*_{\kparb\sigma} \;\theta(k'-\kpar).
\label{betab}
\end{align}
Introducing this coefficient back into Eqs.\ (\ref{Efield}) and (\ref{Power}), we obtain a maximum ratio of the electric field intensity to the incident power
\begin{align}
{\rm max}\left\{\frac{|\Eb(0)|^2}{\mathcal{P}^{\rm inc}}\right\}=\frac{2\sqrt{\epsilonh}\,\omega^2}{3c^3}\,\mathcal{F},
\nonumber
\end{align}
where
\begin{align}
\mathcal{F}=\frac{3}{4\pi k'}\sum_{\sigma={\rm s,p}} \int \frac{d^2 \kparb}{k'_z} \left|\nn\cdot\bb_{\kparb\sigma}\right|^2\theta(k'-\kpar).
\label{FF}
\end{align}
In the absence of a surface ($\tilde{r}_{\kpar\sigma}=0$), $\mathcal{F}=1$ and we recover the result of Sec.\ \ref{maximalhomo}. For a perfect-conductor surface and $a=0$ (i.e., $\trp=1$ and $\trs=-1$), the maximal focal intensity is increased by a factor $\mathcal{F}=4$, independently of the orientation of $\nn$. For real materials and finite separations, $\mathcal{F}$ can still take high values (see supplementary Fig.\ \ref{FigS1}), from which we anticipate an important effect when a particle is added to the system.

\subsection{Maximum coupling to surface polaritons by a small particle}
\label{Maxcoupling}

In order to investigate maximal coupling to surface modes mediated by the presence of the particle in the configuration of Fig.\ \ref{Fig1}(a), we find it convenient to first study the power scattered by the dipole $\pb$ induced at the particle. The total power emanating from that dipole can be obtained by integrating the radial Poynting vector over a small sphere surrounding it, which yields \cite{paper053,NH06} $\mathcal{P}^{\rm scat}=2\omega\left[2k'^3|\pb|^2/3\epsilonh+{\rm Im}\{\pb^*\cdot\Eb^{\rm ind}(0)\}\right]$, where $\Eb^{\rm ind}(0)=\mathcal{G}\cdot \pb$ is the electric field induced by the dipole at the particle position. More precisely,
\begin{align}
\mathcal{P}^{\rm scat}=2\omega\left[\frac{2k'^3|\pb|^2}{3\epsilonh}+|\pb_\parallel|^2{\rm Im}\{\mathcal{G}_\parallel\}+|p_z|^2{\rm Im}\{\mathcal{G}_\perp\}\right].
\label{Pabs}
\end{align}
The first term in this expression corresponds to the radiated power in an infinite host medium of permittivity $\epsilonh$, whereas the second and third terms are proportional to the surface reflectivity through $\mathcal{G}$, which receives contributions from all parallel wave vectors [see Eq.\ (\ref{G})], including $\kpar>k'$ outside the light cone. Here, we are interested in coupling to surface modes, and therefore, we consider the partial contribution to $\mathcal{P}^{\rm scat}$ arising from a certain wave vector range $(k_{\parallel1},k_{\parallel2})$ outside the light cone, characterized by a transferred power
\begin{align}
\mathcal{P}^{\rm surf}=2\omega\left[|\pb_\parallel|^2{\rm Im}\{\mathcal{G}^{\rm surf}_\parallel\}+|p_z|^2{\rm Im}\{\mathcal{G}^{\rm surf}_\perp\}\right],
\label{Psurf}
\end{align}
where $\mathcal{G}^{\rm surf}_s$ with $s=\parallel,\perp$ is given by Eq.\ (\ref{G}) by limiting the integrals to the $k_{\parallel1}<\kpar<k_{\parallel2}$ range. For semi-infinite lossy media and films placed in a symmetric dielectric environment, this part of the scattered power must necessarily be absorbed by the surface (see below). Following a similar procedure as in Sec.\ \ref{maximalhomo} (see Appendix\ \ref{derivationPsurf} for details), we find that $\mathcal{P}^{\rm surf}$ is maximized when the incident beam coefficients are given by Eq.\ (\ref{betab}). We thus conclude that the same choice of beam profile remarkably produces both a maximum field at the position of the particle (i.e., at a distance $a$ from the surface) as a function of such profile and a maximum power transfer to the surface, irrespective of the choice of $(k_{\parallel1},k_{\parallel2})$ and the polarizatiblity of the particle. Incidentally, the maximum field as a function of spatial position is generally displaced with respect to the particle location.

Assuming an optimum incident beam profile as given by Eq.\ (\ref{betab}), we still have more degrees of freedom to maximize the coupling to the surface: the separation $a$ and the particle polarizability $\aE$. From Eqs.\ (\ref{pbsurf}) and (\ref{Psurf}), we find that, for particles with polarization either parallel ($s=\parallel$) or perpendicular ($s=\perp$) to the surface, $\aE$ enters $\mathcal{P}^{\rm surf}$ through an overall factor $|1/\alpha_{{\rm E}s}-\mathcal{G}_s|^{-2}$, and therefore, surface coupling is maximized by minimizing $|\alpha_{{\rm E}s}-\mathcal{G}_s|$. Additionally, as noted above, the optical theorem imposes \cite{V1981} ${\rm Im}\{1/\alpha_{{\rm E}s}\}\le-2k'^3/3\epsilon$, with the equal sign corresponding to nonabsorbing particles. Also, because $\mathcal{P}^{\rm scat}$ cannot be negative, Eq.\ (\ref{Pabs}) imposes the inequality ${\rm Im}\{\mathcal{G}_s\}\ge-2k'^3/3\epsilon$. These general constrains of $\alpha_{\rm E}$ and $\mathcal{G}$ imply ${\rm Im}\{1/\alpha_{{\rm E}s}-\mathcal{G}_s\}\le0$, and we conclude that the coupling is maximized for particles that simultaneously satisfy the conditions
\begin{subequations}
\begin{align}
{\rm Im}\{1/\alpha_{{\rm E}s}\}=-2k'^3/3\epsilon,   \quad & \text{(lossless)}  \label{condition1}\\
{\rm Re}\{1/\alpha_{{\rm E}s}-\mathcal{G}_s\}=0.    \quad & \text{(resonant)}  \label{condition2}
\end{align}
\label{conditions}
\end{subequations}
In particular, Eq.\ (\ref{condition1}) is automatically satisfied if the particle is composed of nonabsorbing materials, whereas Eq.\ (\ref{condition2}) can be approximately fulfilled for feasible scatterers with a vertical size smaller than their separation from the surface (see Sec.\ \ref{realistic}). Both conditions are also met by two-level atoms operated at resonance (see Sec.\ \ref{Rbatom}).

In what follows, we consider films that are either optically thick or placed in a symmetric dielectric environment, so all energy coupled to the surface stays trapped in it. We also consider lossless, resonant particles satisfying Eqs.\ (\ref{conditions}), as well as optimized beam profiles given by Eq.\ (\ref{betab}). For simplicity, we discuss separately particles with their polarizability oriented either parallel ($s=\parallel$ and $\pb\parallel\zz$) or perpendicular ($s=\perp$ and $\pb\parallel\xx$) to the surface. Under these conditions, putting the above expressions together, the maximum fraction of light power coupled to surface modes outside the light cone is (see details in Appendix\ \ref{derivationall})
\begin{equation}
\mathcal{A}_{\rm max}={\rm max}\left\{\frac{\mathcal{P}^{\rm surf}}{\mathcal{P}^{\rm inc}}\right\}=\frac{2 g_s^{\rm surf} \mathcal{F}_s}{(1+g_s)^2},
\label{eq:Abs}
\end{equation}
where
\begin{subequations}
\begin{align}
g_{\parallel} &= \frac{3}{4} \int_0^\infty q dq\; {\rm Re}\left\{\frac{\tilde{r}_{\kpar{\rm s}} - (1-q^2)\tilde{r}_{\kpar{\rm p}}}{\sqrt{1-q^2+\ii0^+}}\right\}, \label{eq:gpar} \\ 
g_{\perp} &= \frac{3}{2} \int_0^\infty q^3 dq\; {\rm Re}\left\{\frac{\tilde{r}_{\kpar{\rm p}}}{\sqrt{1-q^2+\ii0^+}}\right\}, \label{eq:gperp} \\ 
\mathcal{F}_{\parallel} &= \frac{3}{4} \int_0^1 \frac{q dq}{\sqrt{1-q^2+\ii0^+}} \label{eq:hpar}\\
&\quad\quad\quad\quad\times\left[ |1+\tilde{r}_{\kpar{\rm s}}|^2 + (1-q^2) |1-\tilde{r}_{\kpar{\rm p}}|^2 \right],  \label{eq:hpara}\\ 
\mathcal{F}_{\perp} &= \frac{3}{2} \int_0^1 \frac{q^3 dq}{\sqrt{1-q^2+\ii0^+}}  |1+\tilde{r}_{\kpar{\rm p}}|^2,  \label{eq:hperp}
\end{align}
\label{allequation}
\end{subequations}
$g_s^{\rm surf}$ is the same as $g_s$ with $q$ integrated from 1 instead of 0, the integration variable is $q=\kpar/k'$, the square roots are taken to yield positive real parts, and we have chosen $k_{\parallel1}=k'$ and $k_{\parallel1}=\infty$ (i.e., the entire range outside the light cone). Also, $\mathcal{F}_s$ in Eqs.\ (\ref{eq:hpara}) and (\ref{eq:hperp}) coincides with the maximum focal field intensity enhancement at a distance $a$ from the surface relative to the intensity in an infinite homogeneous medium, as given by Eq.\ (\ref{FF}), where we now specify the field orientation through $s$.

\section{Results and discussion}

For simplicity, we consider in what follows semi-infinite lossy surfaces and films placed in a symmetric dielectric environment, although we note that part of the power fraction in Eq.\ (\ref{eq:Abs}) could be released as radiation through the far interface of thin films if the permittivity of the dielectric in that region exceeds the one in the near side. Additionally, we evaluate Eqs.\ (\ref{eq:Abs})-(\ref{allequation}) for different types of films and materials (described as explained in Appendix\ \ref{app:materials}) and find the optimum distance $a$ at which an absolute value of $\mathcal{A}_{\rm max}$ is achieved as a function of light frequency.

\begin{figure*}[ht]
\centering
\includegraphics[width=1.0\textwidth]{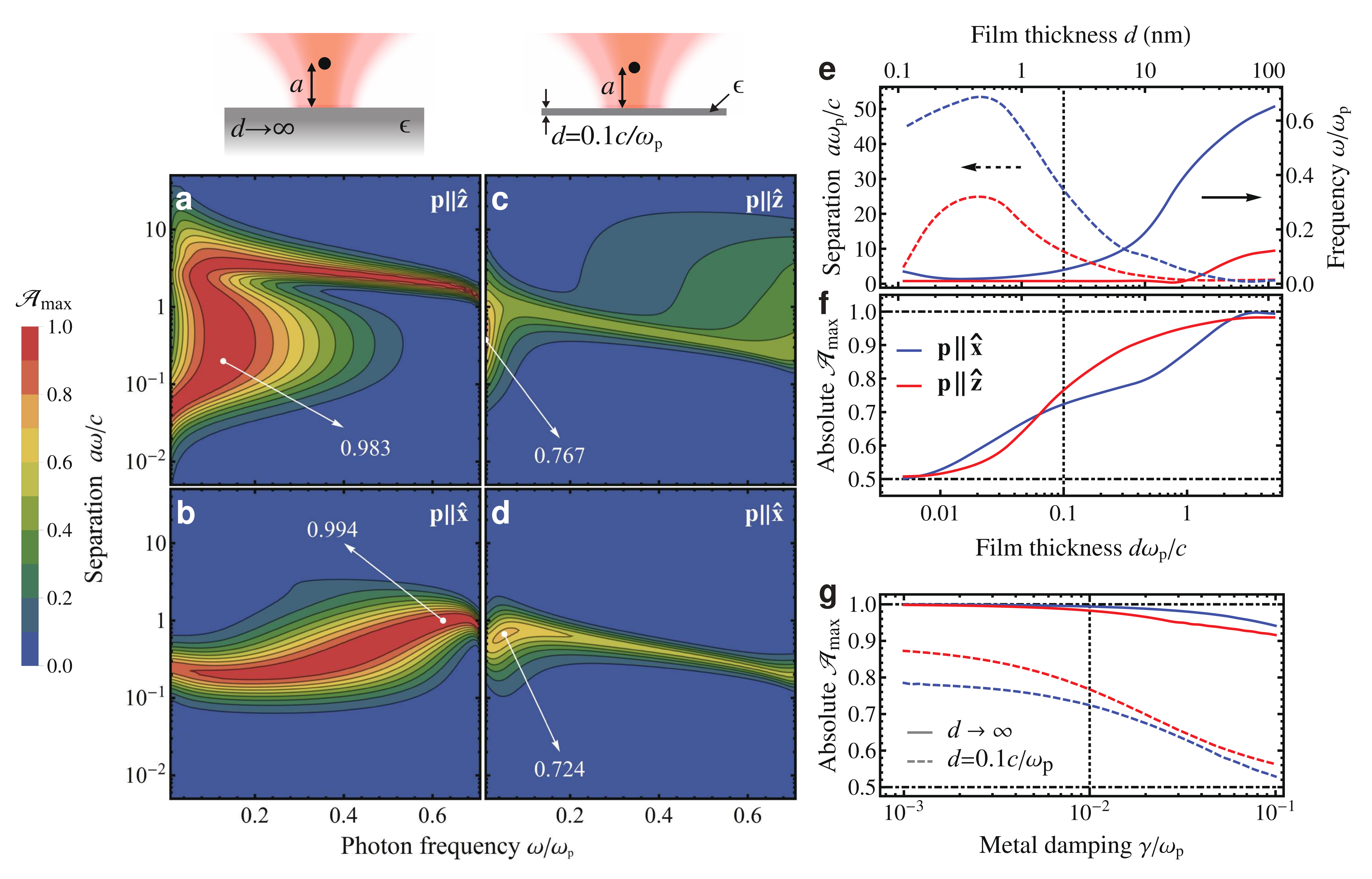}
\caption{Optimum coupling to plasmons in Drude metal films. We study the maximum coupling that can be achieved under the configuration of Fig.\ \ref{Fig1}(a) to surface plasmons in films described through a permittivity $\epsilon(\omega)=1-\wp^2/\omega(\omega+\ii\gamma)$, as calculated for optimized beam profiles depending on frequency, orientation of the particle polarization, and geometrical parameters. (a-d) Dependence of the maximum coupling fraction $\mathcal{A}_{\rm max}$ on frequency $\omega$ and particle-surface separation $a$ for infinite (a,b) and finite (c,d) film thickness $d$ with the particle polarizability directed along either $\zz$ (a,c) or $\xx$ (b,d). (e) Frequency and particle-surface separation for which an absolute maximum coupling is achieved as a function of film thickness. (f) Absolute maximum coupling under the conditions of (e). (g) Drude damping dependence of the absolute maximum coupling for thick and thin films. In (a-f) we take a Drude damping $\gamma=0.01\,\wp$. In (e-g), we show results for both $\zz$ (red curves) and $\xx$ (blue curves) particle polarization. In all plots, we consider self-standing films and normalize $a$ to $c/\omega$, $d$ to $c/\wp$, and $\omega$ and $\gamma$ to $\wp$.}
\label{Fig2}
\end{figure*}

\subsection{Coupling to plasmons in Drude metal films}

We study light coupling to plasmons by considering self-standing metallic films described by the Drude permittivity $\epsilon(\omega)=1-\wp^2/\omega(\omega+\ii \gamma)$. The optimum light coupling $\mathcal{A}_{\rm max}$ to plasmons supported by these films is presented as a function of light frequency and particle-surface separation for thick and thin films in Fig.\ \ref{Fig2}(a-d). These plots exhibit sharp features that define an optimum particle-surface separation $a$ for each light frequency $\omega$, which we attribute to the dominant effect of coupling to the metal surface plasmons. Large coupling nearing 100\% is observed for both in-plane and out-of-plane orientations of the particle polarizability in thick films, while $\mathcal{A}_{\rm max}$ still takes substantial values for thin films. For reference, the Drude model provides a good description of ultrathin silver and gold films \cite{paper300} with $\hbar\wp\sim9\,$eV, so for these materials we have $d=0.1\,c/\wp\approx2.2\,$nm in Fig.\ \ref{Fig2}(c,d), a thickness for which plasmons have been experimentally observed in crystalline silver samples \cite{paper335}. Comparing in-plane and out-of-plane particle polarization, we find that $\mathcal{A}_{\rm max}$ presents a more complex dependence on $a$ and $\omega$ for $\pb\parallel\zz$, and in particular, large coupling $>90\%$ is observed for thick films at $\omega\sim0.15\,\wp$ over a wide range of particle positions.

Now, for each film thickness $d$, we search for the values of $a$ and $\omega$ at which $\mathcal{A}_{\rm max}$ reaches an absolute maximum as a function of those parameters. The corresponding optimum values of $a$ and $\omega$ present a strong dependence on $d$ [Fig.\ \ref{Fig2}(e)], with thicker films producing larger absolute $\mathcal{A}_{\rm max}$ [Fig.\ \ref{Fig2}(f)], although this quantity always stays above $50\%$ regardless of how thin the film is (see $\omega$-$a$ maps of $\mathcal{A}_{\rm max}$ for different film thicknesses in supplementary Fig.\ \ref{FigS2}). Incidentally, when moving toward the $d\rightarrow0$ limit, we find that the depletion of material is compensated by incident beam profiles that are strongly weighted near grazing incidence (see supplementary Fig.\ \ref{FigS3}). The metal damping rate $\gamma$ also plays a significant role on the absolute $\mathcal{A}_{\rm max}$ [Fig.\ \ref{Fig2}(g)], particularly for thin films. Interestingly, $\xx$ polarization is slightly more efficient in the semi-infinite metal limit with independence of the choice of $\gamma$, while the opposite is true for thin films.

We note that coupling can be improved substantially in asymmetric environments with light incident from the high-index side (see supplementary Fig.\ \ref{FigS4}), reaching nearly 100\% for a thin film like that in Fig.\ \ref{Fig2}(c,d), but now supported in Si ($\epsilon=12$), which is an experimentally feasible configuration \cite{paper331}.

We also remark that coupling to surface modes is indeed dominated by plasmons, particularly for $\zz$ dipole orientation, as we show in supplementary Fig.\ \ref{FigS5} by plotting the integrand of Eqs.\ (\ref{eq:gpar}) and (\ref{eq:gperp}) under optimum coupling conditions as a function of $q$ and $\omega$, although nonresonant absorption can also play a substantial role in thick metals for $\xx$ dipole orientation.

\begin{figure*}[ht]
\centering
\includegraphics[width=1.0\textwidth]{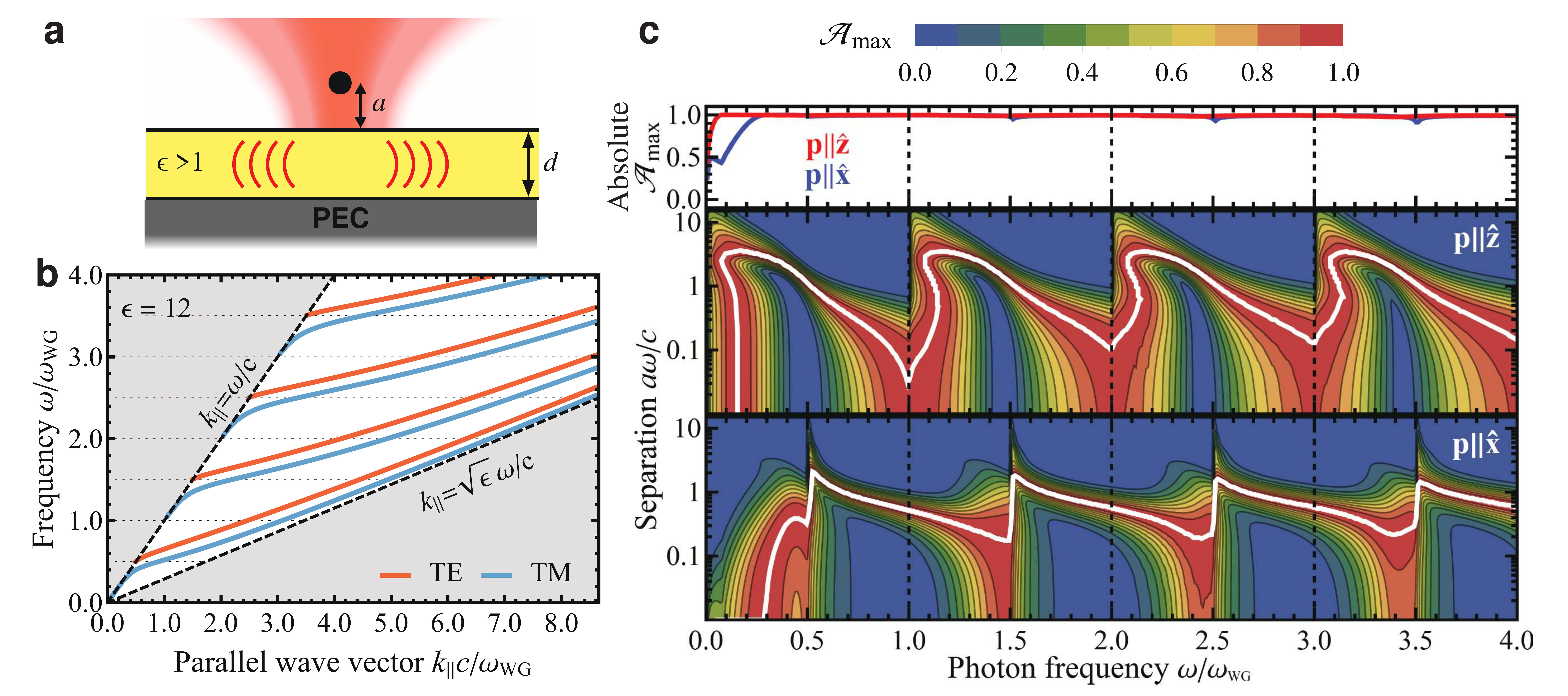}
\caption{Complete coupling to dielectric waveguide modes. (a) Scheme of the system under consideration, comprising a planar dielectric waveguide of thickness $d$ and permittivity $\epsilon$, supported on a perfect-electric-conductor (PEC) substrate, and including a neighboring dipolar scatterer at a distance $a$. (b) Dispersion relation of the TE- and TM-polarized waveguide modes supported by this structure, which are confined in the $\kpar/\sqrt{\epsilon}<\omega/c<\kpar$ region. The axes are normalized using $\omega_{\rm WG}=(\pi c/d)/\sqrt{\epsilon-1}$. (c) Optimized light-to-waveguide coupling maps as a function of light frequency and particle-waveguide separation for two different dipole orientations, as indicated by labels. The white contours correspond to $\mathcal{A}_{\rm max}=1$ (100\% coupling). Panels (b) and (c) correspond to $\epsilon=12$ and are universal for any film thickness $d$.}
\label{Fig3}
\end{figure*}

\subsection{Coupling to dielectric waveguide modes}

High-index planar films provide an attractive approach to guide light because they present comparatively small losses. We consider a film of thickness $d$ and permittivity $\epsilon$ supported on a perfect-electric-conductor (PEC) substrate [Fig.\ \ref{Fig3}(a)], for which waveguide modes of TE and TM symmetry are confined to the $k<\kpar<k\sqrt{\epsilon}$ region, as shown in Fig.\ \ref{Fig3}(b). (We present analogous results for self-standing films in supplementary Fig.\ \ref{FigS6}.) Following the methods discussed in Appendix\ \ref{waveguides}, we can efficiently evaluate Eqs.\ (\ref{eq:Abs}) and (\ref{allequation}) to yield the coupling maps plotted in Fig.\ \ref{Fig3}(c). These plots, which correspond to $\epsilon=12$ (e.g., Si in the near-infrared spectral region), are universal for any arbitrary film thickness $d$, as this parameter is embedded in the characteristic frequency $\omega_{\rm WG}=(\pi c/d)/\sqrt{\epsilon-1}$ (see Appendix\ \ref{waveguides}). The dependence of $\mathcal{A}_{\rm max}$ on separation $a$ and frequency $\omega$ is dominated by the dispersion relation of the modes, and in particular, successive modes emerge at periodic intervals determined by $\omega_{\rm WG}$ as $\omega$ increases. The emergence of every new TE or TM mode produces a sharp feature in the $\omega$-$a$ dependence of the coupling strength for $\xx$ or $\zz$ particle polarization, respectively. Although the overall absorption varies slightly with the orientation of the particle dipole, coupling efficiencies $>90\%$ are observed over broad ranges of the separation $a$ at nearly all frequencies, and complete coupling is also found at frequency-dependent optimum values of $a$. Interestingly, as the number of modes increases with frequency, each of them must receive a smaller fraction of energy from the incident light, which is directly related to the $j$ sums in the expressions for $g_s^{\rm surf}$ presented in Appendix\ \ref{waveguides}. Conversely, at low frequencies below $0.5\,\omega_{\rm WG}$, there is only one waveguide mode, which has TM symmetry, and for which 100\% coupling can be realized when the particle polarizability oriented along $\zz$. Further examination of the partial contribution of different modes to the overall coupling (supplementary Fig.\ \ref{FigS7}) reveals that emerging modes of increasingly higher order become dominant as the frequency increases (i.e., the mode that is closest to the light cone for a given frequency generally dominates the coupling of light to the waveguide), and more precisely, for out-of-plane particle dipole orientation, we encounter alternating frequency intervals in which a single mode accounts for nearly 100\% coupling. Incidentally, coupling to self-standing Si waveguides is reduced compared to PEC-supported films, but the absolute maximum is still $>70\%$ (see supplementary Fig.\ \ref{FigS6}).

\begin{figure*}[ht]
\centering
\includegraphics[width=0.8\textwidth]{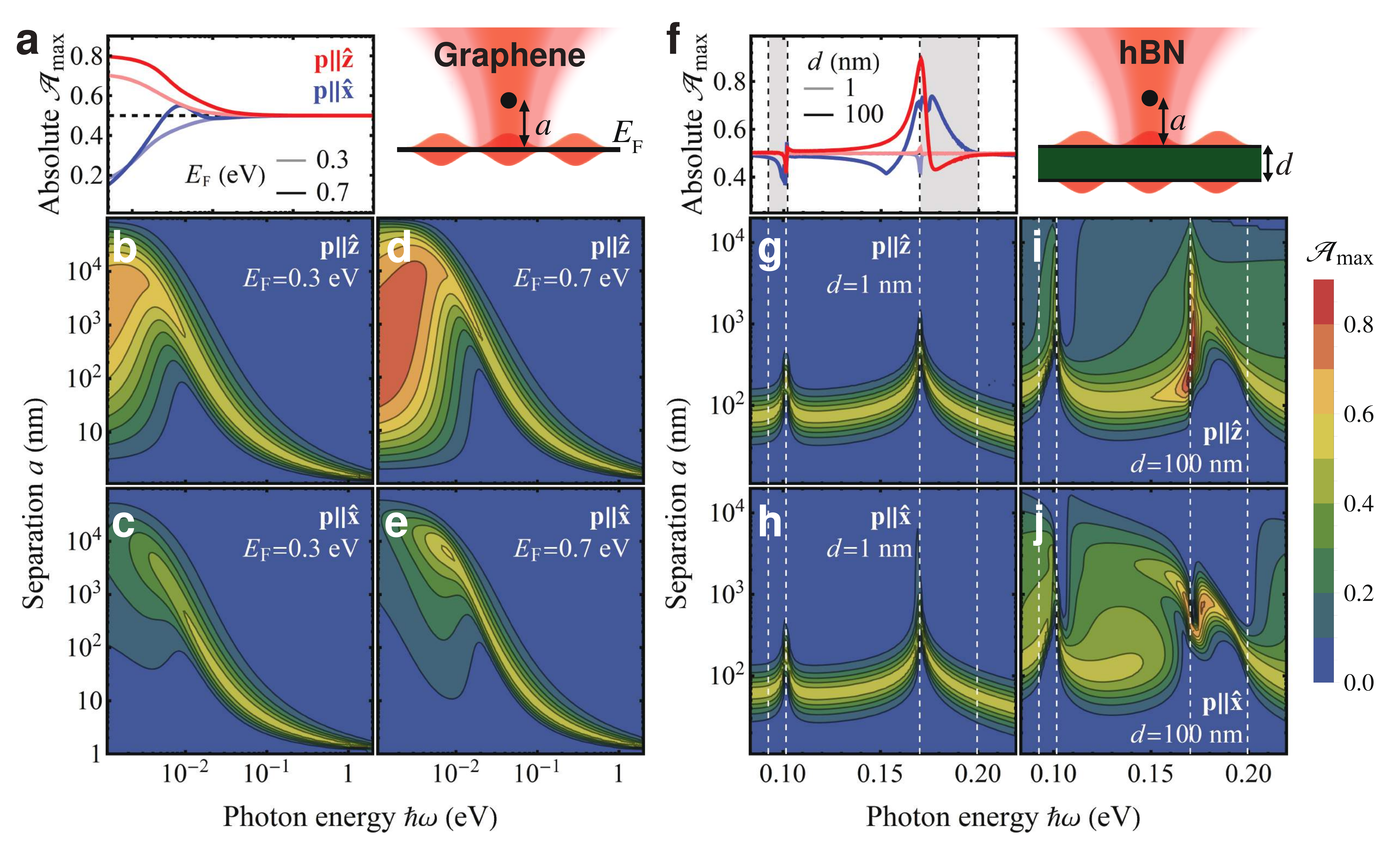}
\caption{Optimum coupling to polaritons in 2D materials. (a-e) Coupling to graphene plasmons for two different values of the doping Fermi energy $\EF$. We consider $\zz$ and $\xx$ dipole orientations, as indicated by labels. Plots in (b-e) show the dependence of the coupling fraction $\mathcal{A}_{\rm max}$ on photon energy and dipole-surface separation. Panel (a) shows the maximum coupling for an optimum separation as a function of photon energy. (f-j) Same as (a-e) for hBN films of different thickness $d$ (see labels). Vertical dashed lines indicate the limits of the hBN Reststrahlen bands.}
\label{Fig4}
\end{figure*}

\subsection{Coupling to surface polaritons in 2D materials: Graphene and hBN}

We now move to the optimization of light coupling to 2D polaritons supported by thin films of 2D materials, and specifically, we concentrate on graphene and hBN as prototypical examples of plasmonic and hyperbolic-polaritonic van der Waals materials \cite{paper283}. For simplicity, we consider self-standing films.

The results that we obtain for graphene [Fig.\ \ref{Fig4}(a-e)] are analogous to those for coupling to thin-metal-film plasmons [Fig.\ \ref{Fig2}(c,d)] because the carbon monolayer material can also be well described as a Drude metal with a bulk plasma frequency \cite{paper254} $\wp=(2e\hbar)\sqrt{\EF/d}$, where $d\approx0.33\,$nm is the atomic layer thickness estimated from the interatomic plane distance in graphite. Remarkably, the optimized photon-to-plasmon coupling efficiency reaches or exceeds $50\%$ over a wide spectral range, in contrast to the much poorer maximum coupling previously predicted when the scattering particle is close to the graphene surface \cite{paper331}. At low light frequencies, the coupling efficiency reaches even higher values approaching $80\%$ for a particle with $\zz$ polarization situated at distances of 100s to 1000s of nm from the graphene plane [Fig.\ \ref{Fig4}(a,d)]. Incidentally, we assume a Drude damping $\hbar\gamma=2\,$meV, so coupling at very low frequencies $\omega\lesssim\gamma$ can be contributed by direct inelastic absorption, rather than plasmon creation, but nevertheless, the features observed in Fig.\ \ref{Fig4}(b-e) are dominated by plasmons (see supplementary Fig.\ \ref{FigS5}). Also, an increase in graphene doping generally enhances light coupling to plasmons [cf. Fig.\ \ref{Fig4}(b-c) and Fig.\ \ref{Fig4}(d-e)], as doping plays an effect similar to thickness in metal Drude films (see Fig.\ \ref{Fig2}), bringing the plasmon dispersion relation closer to the light line and reducing the photon-plasmon momentum mismatch.

Coupling to hBN films exhibits a rather different behavior because this material only supports phonon polaritons in the two narrow spectral regions known as Reststrahlen bands, which are indicated by vertical dashed lines in Figs.\ \ref{Fig4}(f-j). These bands are distinctly visible in the optimized absorption maps. Outside them, absorption still reaches a maximum of $\sim50\%$ under optimized particle and beam-shape conditions as a result of copuling to regular dielectric waveguide modes similar to those of Fig.\ \ref{Fig3} (i.e., the material behaves as a normal dielectric). Now, although coupling to surface propagating modes only slighly increases absorption above that background level for 1\,nm film thickness, we find that thicker hBN films present a strong enhancement inside the Reststrahlen bands, driven by excitation of phonon polaritons and producing an absorption as high as $>80\%$ for a 100\,nm film.

\begin{figure*}[ht]
\centering
\includegraphics[width=0.90\textwidth]{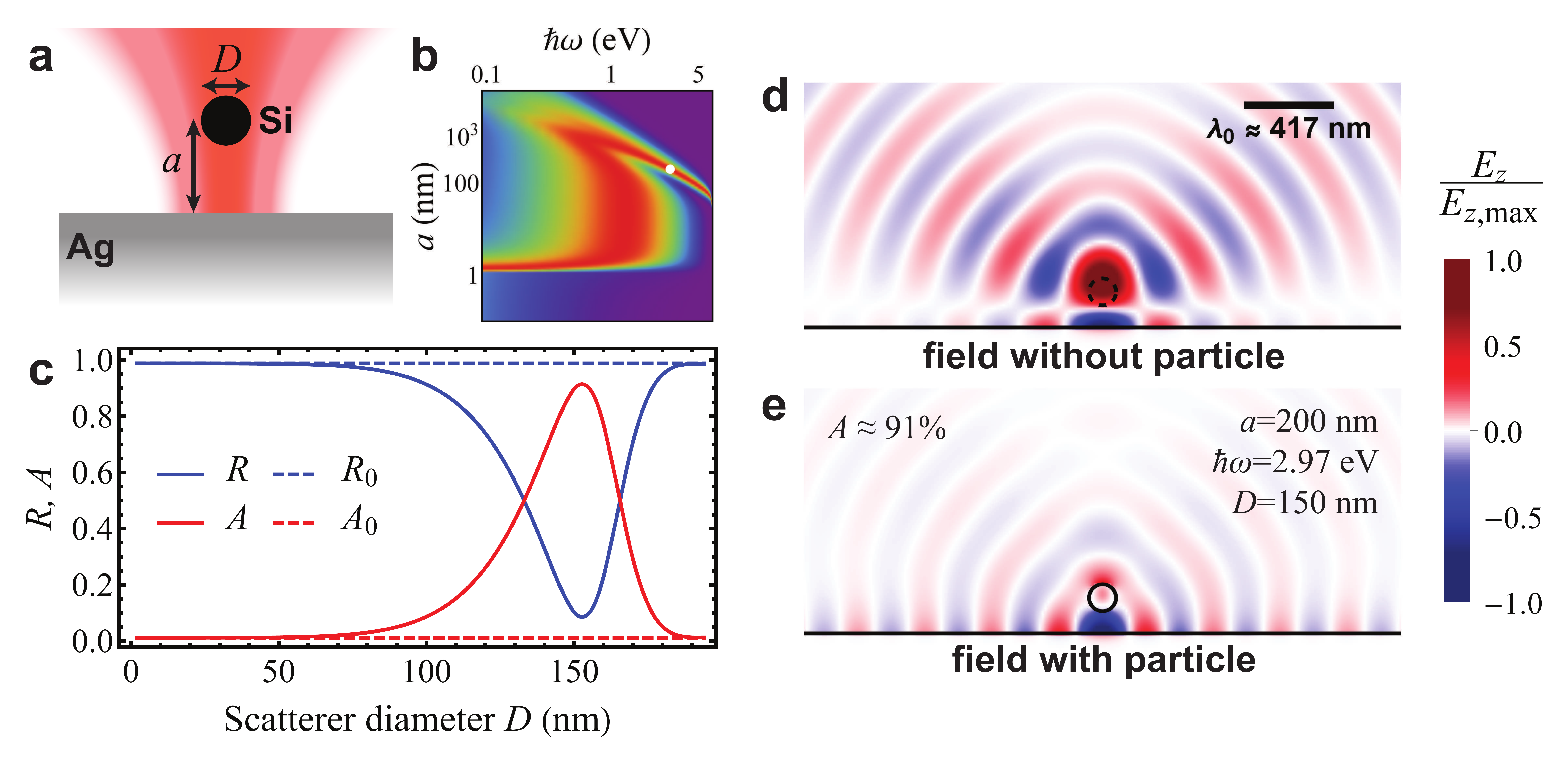}
\caption{Optimized absorption for finite-size scatterers. (a) System under consideration, consisting of a silicon sphere ($\epsilon=12$) of diameter $D$ placed in front of a semi-infinite silver surface. (b) Optimized absorption as a function of particle-surface separation $a$ and photon energy $\hbar\omega$. (c) Numerically simulated reflection ($R,R_0$) and absorption ($A,A_0$) for the optimized incident field in the absence ($R_0,A_0$) and presence ($R,A$) of the silicon sphere as a function of diameter $D$. (d,e) Out-of-plane electric field amplitude distribution corresponding to the optimized field for the separation and frequency indicated by a white dot in (b) [see parameters indicated by labels in (e)]. We present numerical simulations in the absence (d) and presence (e) of the particle, with the latter represented by open and solid circles, respectively, the silver surface indicated by horizontal solid lines, and the wavelength shown by the scale bar in (d).}
\label{Fig5}
\end{figure*}

\subsection{Dielectric particle as a perfect scatterer}
\label{realistic}

The above results assume the existence of lossless resonant scatterers that can be placed at optimized distances from the surface. As a practical example of such type of perfect scatterer, we consider high-index spherical particles, which host strong dipolar Mie resonances of electric and magnetic character that satisfy the conditions of Eqs.\ (\ref{conditions}) to a good approximation (see supplementary Fig.\ \ref{FigS8}). The spectral positions of these resonances depend on the permittivity $\epsilon$ and diameter $D$ of the particle, so these parameters allow us to place the scatterer resonance at the desired frequency for which we intend to optimize absorption. We illustrate this concept by considering a silicon sphere ($\epsilon=12$) placed above a semi-infinite silver surface [see Fig.\ \ref{Fig5}(a)] and choose to orient the induced particle dipole along the $z$ direction using a p-polarized incident beam. We start by computing the separation- and frequency-dependent map of optimized absorption by the metal [Fig.\ \ref{Fig5}(b)] assuming a lossless resonant point particle. This map presents an absolute maximum at $a\approx 200\,$nm and $\hbar\omega \approx2.97\,$eV when using an incident beam profile as prescribed by Eq.\ (\ref{betab}) and graphically plotted in Fig.\ \ref{Fig1}(b). We now maintain these parameters fixed (i.e., $a$, $\omega$, and the beam profile), but substitute the point dipole by a finite-size silicon particle and numerically simulate the absorption $A$ and reflection $R$ of this system using a finite-difference frequency-domain method. The resulting dependence of $A$ and $R$ on particle diameter is shown in Fig.\ \ref{Fig5}(c) in the presence (solid curves) and absence (dashed curves) of the particle. Without the particle, $99\%$ of the incident light is reflected. However, the presence of a silicon sphere produces a resonant absorption feature that reaches $A>90\%$ for a diameter $D\approx 150\,$nm. The corresponding electric near-field amplitude distribution further illustrates almost complete reflection in the absence of the particle [Fig.\ \ref{Fig5}(d)] and strong depletion of reflection when the resonant particle is present [Fig.\ \ref{Fig5}(e)]. We note that, for the resonant particle diameter, a self-standing Si sphere hosts an electric dipole Mie mode at the operating photon energy $\hbar\omega\approx 2.97\,$eV (see supplementary Fig.\ \ref{FigS8}). These numerical results confirm that high-index dielectric particles constitute good candidates of lossless resonant absorbers for practical implementations in the present context.

\subsection{Two-level atomic scatterers}
\label{Rbatom}

In order to evaluate the feasibility of using atomic scatterers to satisfy the conditions imposed by Eqs.\ (\ref{conditions}), we consider a two-level optical resonance of an atom or molecule embedded in a host medium of permittivity $\epsilon$ (e.g., $\epsilon=1$ for optically-trapped atoms in vacuum or $\epsilon>1$ for molecules trapped in a host dielectric under cryogenic conditions that preserve the two-level character of the system \cite{RWL12}). Because the interior of the atom or molecule is not permeated by the host material, assuming that it occupies a small spherical void, the resonance transition dipole needs to be corrected by a depolarization factor $f_\epsilon=9\epsilon^2/(2\epsilon+1)^2$ \cite{YGB1988}, which affects the radiative decay rate $\Gamma=\Gamma_0f_\epsilon\sqrt{\epsilon}$ relative to the one in vacuum, $\Gamma_0$. Near a lossless two-level optical resonance of frequency $\omega_0$, we can approximate the atomic polarizability as \cite{L1983,VST96}
\begin{equation}
\aE(\omega)\approx\frac{3c^3}{2\sqrt{\epsilon}\,\omega_0^2}\frac{\Gamma}{\omega_0^2-\omega^2-\ii\Gamma\omega^3/\omega_0^2},
\label{eq:alpha:atom}
\end{equation}
which directly satisfies the lossless condition (\ref{condition1}). The resonance condition (\ref{condition2}), which requires large values of ${\rm Re}\{1/\aE\}$ to compensate ${\rm Re}\{\mathcal{G}\}$, can be fulfilled for $\omega$ near the resonance. In particular, we show that this is feasible for Rb atoms by exploiting their 750\,nm resonance line ($\Gamma\approx25\,$MHz) \cite{SWC03}. Complete light-to-waveguide coupling can be realized by using a single Rb atom optically-trapped in vacuum at a distance of 100\,nm above a 137-nm-thick silicon waveguide under the configuration of Fig.\ \ref{Fig3} (see supplementary Fig.\ \ref{FigS9}); the resonance conditions are met using TM illumination with $\sim1.5\,$MHz red detuning.

\section{Conclusions}

In summary, we have demonstrated based on rigorous electromagnetic analytical theory that a small particle placed in front of a polariton-supporting planar surface can boost the coupling of an external focused light beam to the polaritons propagating in the material. Coupling can be complete for waveguide modes in dielectric films, and it also reaches near-unity values for plasmonic materials such as silver and graphene. The conditions for optimum coupling require a suitable particle-surface distance and a shaped light beam profile, for which we offer detailed general prescriptions in closed-form analytical expressions, depending on the operation light frequency and the surface material and structure. Additionally, the particle must be lossless and resonant, which are feasible conditions met by high-index Mie scatterers and two-level atoms or molecules. In a practical configuration for the near-infrared spectral range, the particle could consist of lithographically carved Si or Ge particles (e.g., spheres or disks), fixed at a designated distance from the surface by an embedding lower-index host medium.

The present results focus on scatterers dominated by an electric dipole resonance, but our analysis can be straightforwardly extended to magnetic dipolar scatterers, for which the beam profile has a different optimum profile, as we illustrate for an isolated particle placed in an infinite homogeneous medium. As an interesting extension of this work, optimum beam profiles could be found for particles having an arbitrary multipolar polarizability, for which we conjecture complete coupling to surface polaritons if they are also made of nonabsorbing materials and exhibit a sufficiently strong optical resonance. While all of these ideas refer to systems operating at a single light frequency, large coupling within an extended frequency range should be possible using those particles if they exhibit multiple resonances within that range.

Our study offers a practical route toward the design of localized surface polariton sources at predetermined positions, fed by external shaped light beams and integrated in nanophotonic devices that could be based on planar surfaces and engineered scatterers placed in front of the intended source locations. The same principles apply to two-level atoms, which could be optically trapped at distances $\sim100\,$nm from the surface and serve as mediators of light-polariton coupling as we demonstrate here.

\section*{Acknowledgments} 

This work has been supported in part by the Spanish MINECO (MAT2017-88492-R and SEV2015-0522), ERC (Advanced Grant 789104-eNANO), the Catalan CERCA Program, and Fundaci\'{o} Privada Cellex. E.J.C.D. acknowledges financial support from “la Caixa” (INPhINIT Fellowship Grant 1000110434, LCF/BQ/DI17/11620057) and the EU (Marie Sk\l{}odowska-Curie Grant 713673).

\appendix

\section{Maximization of surface coupling}
\label{derivationPsurf}

We obtain the beam profile that maximizes the ratio $\mathcal{A}=\mathcal{P}^{\rm surf}/\mathcal{P}^{\rm inc}$ by imposing the vanishing of the functional derivative $\delta\mathcal{A}/\delta\beta^{+*}_{\kparb\sigma}=0$ with respect to the incident beam coefficients $\beta^{+*}_{\kparb\sigma}$, or equivalently, $\mathcal{P}^{\rm inc}\,\delta\mathcal{P}^{\rm surf}/\delta\beta^{\nu*}_{\kparb\sigma}=\mathcal{P}^{\rm surf}\,\delta\mathcal{P}^{\rm inc}/\delta\beta^{\nu*}_{\kparb\sigma}$. From the expression of $\mathcal{P}^{\rm inc}=\mathcal{P}^+$ given by Eq.\ (\ref{Power}), we find $\delta\mathcal{P}^{\rm inc}/\delta\beta^{\nu*}_{\kparb\sigma}=[c\sqrt{\epsilonh}/(2\pi)^3k'] k'_z\beta^+_{\kparb\sigma}\propto k'_z\beta^+_{\kparb\sigma}$. Also, from Eq.\ (\ref{Psurf}), we have $\delta\mathcal{P}^{\rm surf}/\delta\beta^{\nu*}_{\kparb\sigma}\propto\pb\cdot\delta\pb^*/\delta\beta^{\nu*}_{\kparb\sigma}$ for a dipole orientation either parallel or perpendicular to the surface, and from Eq.\ (\ref{pbsurf}), $\delta\pb^*/\delta\beta^{\nu*}_{\kparb\sigma}\propto\bb^*_{\kparb\sigma}$. Additionally, by construction $\pb\propto\nn$. Finally, putting these elements together we readily find Eq.\ (\ref{betab}). We note that in this context the symbol $\propto$ involves proportionality constants that are independent of incidence direction (i.e., independent of the parallel wave vector $\kparb$). 

\section{Derivation of Eqs.\ (\ref{eq:Abs}) and (\ref{allequation})}
\label{derivationall}

We consider a particle dipole oriented along either $\nn=\zz$ ($s=\perp$ polarization) or $\nn=\xx$ ($s=\parallel$ polarization), for which the optimum beam profile is given by Eq.\ (\ref{betab}) as $\beta^+_{\kparb\sigma}=(C/k'_z)\nn\cdot\bb^*_{\kparb\sigma}\;\theta(k'-\kpar)$, where $C$ is a $\kparb$-independent constant. Introducing this expression into Eq.\ (\ref{pbsurf}), we identify the same integral as in Eq.\ (\ref{FF}), from which we obtain
\begin{align}
\pb=\nn\frac{C}{1/\alpha_{{\rm E}s}-\mathcal{G}_s}\frac{k'}{3\pi}\mathcal{F}_s.
\label{I2}
\end{align}
Now, using the explicit expressions of $\bb_{\kparb\sigma}$ and $\eh^\nu_{\kparb\sigma}$ given in Eq.\ (\ref{bb}) and Sec.\ \ref{sec2}, we readily derive Eqs.\ (\ref{eq:hpar}) and (\ref{eq:hperp}) for $\mathcal{F}_s$. Additionally, for the lossless, resonant particles under consideration [see Eqs.\ (\ref{conditions})], we can write
\begin{align}
1/\alpha_{{\rm E}s}-\mathcal{G}_s=-\ii(2k'^3/3\epsilonh+{\rm Im}\{\mathcal{G}\}),
\label{I3}
\end{align}
and from Eq.\ (\ref{G}), we find
\begin{align}
{\rm Im}\{\mathcal{G}_s\}=(2k'^3/3\epsilonh)g_s,
\label{I4}
\end{align}
where $g_s$ is defined in Eqs.\ (\ref{eq:gpar}) and (\ref{eq:gperp}). Equations\ (\ref{I2})-(\ref{I4}) allow us to recast Eq.\ (\ref{Psurf}) for the power coupling to the surface as
\begin{align}
\mathcal{P}^{\rm surf}=\frac{\sqrt{\epsilonh}c}{3\pi^2}|C|^2 \frac{g_s^{\rm surf}\mathcal{F}_s^2}{(1+g_s)^2},
\label{Psurff}
\end{align}
where $g_s^{\rm surf}$ is given by Eqs.\ (\ref{eq:gpar}) and (\ref{eq:gperp}) with $q$ integrated in the $(1,\infty)$ range instead of $(0,\infty)$ because the integral involved in the calculation of $\mathcal{G}^{\rm surf}$ [see Eq.\ (\ref{Psurf})] is restricted to $\kpar>k'$. Finally, inserting Eq.\ (\ref{betab}) into Eq.\ (\ref{Power}) and using again Eq.\ (\ref{FF}), we find the incident beam power
\begin{align}
\mathcal{P}^{\rm inc}=\frac{\sqrt{\epsilonh}c}{6\pi^2}|C|^2\,\mathcal{F}_s,
\nonumber
\end{align}
which, combined with Eq.\ (\ref{Psurff}), directly produces Eq.\ (\ref{eq:Abs}).

\section{Material modeling}
\label{app:materials}

We describe the permittivity of silver in the plasmonic and infrared range using the Drude-like model \cite{paper300} $\epsilon_{\rm Ag}(\omega) = \epsilon_{\rm b}-\wp^2/\omega(\omega+\ii\gamma)$ with parameters $\epsilon_{\rm b}=4.0$, $\hbar\wp=9.17\ {\rm eV}$, and $\hbar\gamma=21\ {\rm meV}$ fitted to experimental data \cite{JC1972}.

For graphene, we use the 2D Drude conductivity \cite{GP16} $\sigma_{\rm gr}(\omega)=(\ii e^2\EF/\pi\hbar^2)/(\omega+\ii\gamma)$, which depends on the doping Fermi energy $\EF$ and a phenomenological damping rate here set to $\hbar\gamma=2\,$meV ($\sim330\,$fs lifetime).

We describe hBN films through the in-plane ($s=\parallel$) and out-of-plane ($s=\perp$ along the c-axis) permittivities \cite{CSS07,WLG15}
\begin{equation}
\epsilon_{{\rm hBN},s}(\omega) = \epsilon_{\infty,s} - f_s \frac{\omega_s^2}{\omega(\omega+\ii \gamma_s)-\omega_s^2},
\end{equation}
with $\epsilon_{\infty\parallel}=4.87$, $\epsilon_{\infty\perp}=2.95$, $f_\parallel=1.83$, $f_\perp=0.61$, $\hbar\omega_\parallel=170.1\,$meV, $\hbar\omega_\perp=92.5\,$meV, $\hbar\gamma_\parallel=0.87\,$meV, and $\hbar\gamma_\perp=0.25\,$meV.

For a generic film of isotropic permittivity $\epsilon_2$ and thickness $d$ surrounded by dielectrics of permittivities $\epsilon_1$ and $\epsilon_3$, the reflection coefficients for incidence from the $\epsilon_1$ medium are given by
\begin{align}
r_{\kpar\sigma}&=r_{12}^0 + \frac{t_{12}^0 t_{21}^0 r_{23}^0 \ee^{2\ii k_{z2}d}}{1-r_{21}^0 r_{23}^0 \ee^{2\ii k_{z2}d}},  \label{eq:rp:film}
\end{align}
where $r_{ij}^0=(k_{zi}\eta^{\sigma}_j-k_{zj}\eta^{\sigma}_i)/(k_{zi}\eta^{\sigma}_j+k_{zj}\eta^{\sigma}_i)$ and $t_{ij}^0 = 2 k_{zi}\sqrt{\eta^{\sigma}_i \eta^{\sigma}_j}/(k_{zi}\eta^{\sigma}_j+k_{zj}\eta^{\sigma}_i)$ are the Fresnel reflection and transmission coefficients of the $ij$ interface, $k_{zi}=\sqrt{\epsilon_ik^2-\kpar^2+\ii0^+}$, $\eta^{\rm s}_i=1$, and $\eta^{\rm p}_i=\epsilon_i$. Equation\ (\ref{eq:rp:film}) can be readily applied to silver films by setting $\epsilon_2=\epsilon_{\rm Ag}$. For graphene, the reflection coefficients can be derived from Eq.\ (\ref{eq:rp:film}) by making $\epsilon_2=4\pi\ii\sigma_{\rm gr}/\omega d$ and taking the $d\rightarrow0$ limit, which yields \cite{GP16}
\begin{align}
r_{\kpar{\rm s}}&=\frac{k_{z1}-k_{z2} - 4\pi \sigma_{\rm gr} k /c}{k_{z1}+k_{z2} + 4\pi \sigma_{\rm gr} k /c}, \nonumber \\
r_{\kpar{\rm p}}&=\frac{\epsilon_2 k_{z1}- \epsilon_1 k_{z2} + 4\pi \sigma_{\rm gr} k_{z1}k_{z2} /\omega}{\epsilon_2 k_{z1}+ \epsilon_1 k_{z2} + 4\pi \sigma_{\rm gr} k_{z1}k_{z2} /\omega}. \nonumber
\end{align}
The corresponding reflection coefficients for anisotropic hBN films \cite{DP17} are also given by Eq.\ (\ref{eq:rp:film}) if we redefine $k_{z2}=\sqrt{\epsilon_{\rm hBN}^{\parallel}k^2 - (\epsilon_{\rm hBN}^{\parallel}/\epsilon_{\rm hBN}^{\perp})\kpar^2}$ and take $\epsilon_2(\omega)=\epsilon_{\rm hBN}^{\parallel}$. The sign of the square roots is chosen to yield ${\rm Im}\{k_{zj}\}>0$.

\section{Dielectric waveguides}
\label{waveguides}

\subsection{Pole approximation}
\label{WGgeneral}

Self-standing dielectric waveguides ($\epsilon_1=\epsilon_3=1$, $\epsilon_2=\epsilon$) support TE ($\sigma={\rm s}$) and TM ($\sigma={\rm p}$) guided modes signalled by the vanishing of the denominator in Eq.\ (\ref{eq:rp:film}), leading to the conditions  \cite{J1975}
\begin{subequations}
\begin{align}
\eta_{\sigma} \xi/\xi' &= \tan \left( \xi'/2 \right), \label{eq:WG:1} \\
\eta_{\sigma} \xi/\xi' &= -\cot \left( \xi'/2 \right), \label{eq:WG:2}
\end{align}
\label{WGeqs}
\end{subequations}
where $\xi=d\sqrt{\kpar^2-k^2}$ and $\xi'=d\sqrt{\epsilon k^2-\kpar^2}$ are real in the $k<\kpar<k\sqrt{\epsilon}$ region in which the modes evolve, $\eta_{\rm s}=1$, and $\eta_{\rm p}=\epsilon$. At any frequency $\omega$, we find a finite number $N$ of solutions of Eqs.\ (\ref{WGeqs}) for each polarization, which define the frequency-dependent mode wave vectors $k_{\parallel\sigma j}$ ($j=1,\cdots,N$). Upon inspection, $N$ is simply given by the smallest integer greater than $\omega/\omega_{\rm WG}$, where $\omega_{\rm WG}=(\pi c/d)/\sqrt{\epsilon-1}$ is a characteristic waveguide frequency.

For a waveguide supported on a PEC ($|\epsilon_3|\rightarrow\infty$), such as the one depicted in
Fig.\ \ref{Fig3}(a), Eqs.\ (\ref{WGeqs}) need to be changed to
\begin{subequations}
\begin{align}
\epsilon \xi/\xi' &= \tan \left( \xi' \right),\qquad &\text{for TM modes,} \nonumber \\
\xi/\xi' &= -\cot \left( \xi' \right),\qquad &\text{for TE modes,} \nonumber
\end{align}
\end{subequations}
but the rest of the analysis remains the same as for the self-standing waveguide, except that the number of modes $N$ now depends on polarization and must be defined as the smallest integer greater than $\omega/\omega_{\rm WG}$ and $\omega/\omega_{\rm WG}-1/2$ for TM and TE modes, respectively.

In the vicinity of a mode ($\kpar\approx k_{\parallel\sigma j}$), the film reflection coefficient is well described by the pole approximation \cite{paper331} $r_{\kpar\sigma}\approx\mathcal{R}_{\sigma j}k_{\parallel\sigma j}/(\kpar-k_{\parallel\sigma j}-\ii 0^+)$, where
\begin{align}
\mathcal{R}_{\sigma j}&=\left(k_{\parallel\sigma j}\partial_{\kpar}r^{-1}_{\kpar\sigma}\right)^{-1} \nonumber\\
&=\frac{2\xi\xi'}{(k_{\parallel\sigma j}d)^2} \left[ b \left( \frac{\xi^2 + \xi^{'2} }{\xi \xi'} \right) + \frac{(\eta_{\sigma}\xi)^2+\xi^{'2}}{\eta_{\sigma}\xi'} \right]^{-1}, \nonumber
\end{align}
is a pole residue evaluated by taking the derivative of Eq.\ (\ref{eq:rp:film}) at $\kpar=k_{\parallel\sigma j}$, and we have to set $b=1$ and $b=2$ for PEC-supported and self-standing waveguides, respectively. Now, for low and moderate values of $\omega$, the resonant wave vectors $k_{\kpar\sigma j}$ are well spaced and the reflection coefficient can be approximated as
\begin{align}
r_{\kpar\sigma}\approx\sum_{j=1}^N\frac{\mathcal{R}_{\sigma j}q_{\sigma j}}{q-q_{\sigma j}-\ii 0^+},
\label{rWG}
\end{align}
where $q=\kpar/k$ and $q_{\sigma j}=k_{\parallel\sigma j}/k$.

\subsection{Evaluation of Eqs.\ (\ref{allequation})}

Using Eq.\ (\ref{rWG}) and the identity ${\rm Im}\{1/(x-\ii0^+)\}=\pi\delta(x)$, we find ${\rm Im}\{r_{\kpar\sigma}\}\approx\pi\sum_{j=1}^{N}\mathcal{R}_{\sigma j}q_{\sigma j}\delta(q-q_{\sigma j})$, which allows us to write closed-form analytical expressions for $g_s^{\rm surf}$, defined by Eqs.\, (\ref{eq:gpar}) and (\ref{eq:gperp}) with the lower limit of integration set to $q=1$ instead of 0. More precisely,
\begin{align}
g^{\rm surf}_{\parallel} =& \frac{3\pi}{4} \sum_{j=1}^{N} \bigg(\dfrac{\mathcal{R}_{{\rm s} j} q_{{\rm s} j}^2}{\kappa_{{\rm s} j}} \ee^{-2 k_1 a \kappa_{{\rm s} j}} + \mathcal{R}_{{\rm p} j} q_{{\rm p} j}^2 \kappa_{{\rm p} j} \ee^{-2 k_1 a \kappa_{{\rm p} j}} \bigg), \nonumber \\ 
g^{\rm surf}_{\perp} =& \frac{3\pi}{2} \sum_{j=1}^{N} \dfrac{\mathcal{R}_{{\rm p} j} q_{{\rm p} j}^4}{\kappa_{{\rm p} j}} \ee^{-2 k_1 a \kappa_{{\rm p} j}},
\nonumber
\end{align}
where $\kappa_{\sigma j}=\sqrt{q_{\sigma j}^2-1}$. The remaining integrals in the evaluation of $g_s-g^{\rm surf}_s$ and $\mathcal{F}_s$ only involve well-behaved functions of $q$ in the $(0,1)$ range, so we carry them out numerically.

\pagebreak


\begin{figure*}
\centering
\includegraphics[width=0.9\textwidth]{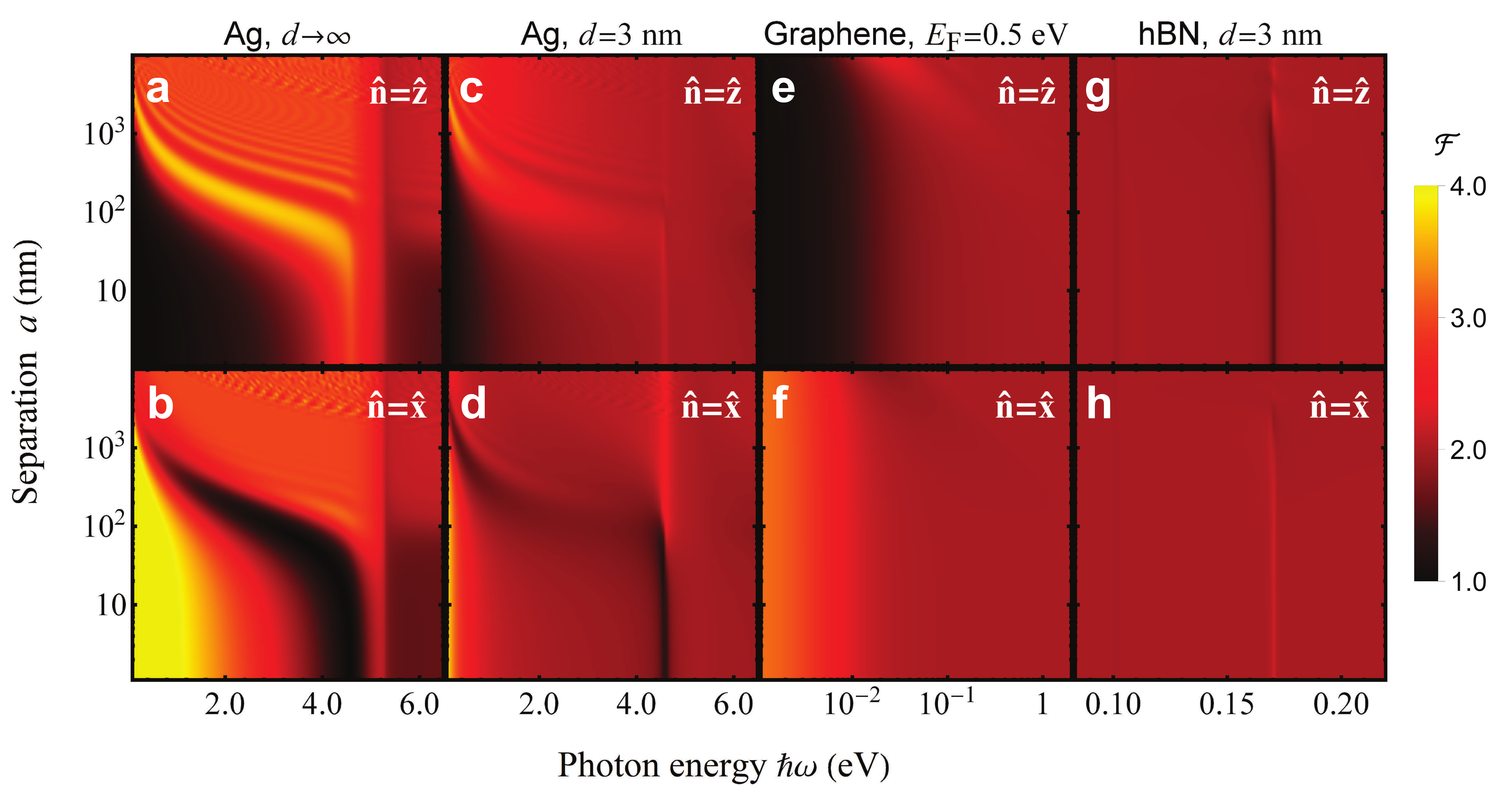}
\caption{{\bf Enhancement of the focal field intensity produced by reflection on a surface.} We plot the optimum enhancement of the field intensity at a distance $a$ from various types of surfaces considered in the main text (see upper labels) as a function of light frequency. The enhancement factor $\mathcal{F}$ is defined relative to the result obtained in the absence of any surface [see Eq.\ (\ref{FF})]. We consider maximization of either out-of-plane (upper plots) or in-plane (lower plots) electric field components. Each point on these plots requires a different optimum beam profile, as prescribed by Eq.\ (\ref{betab}).}
\label{FigS1}
\end{figure*}

\begin{figure*}
\centering
\includegraphics[width=0.9\textwidth]{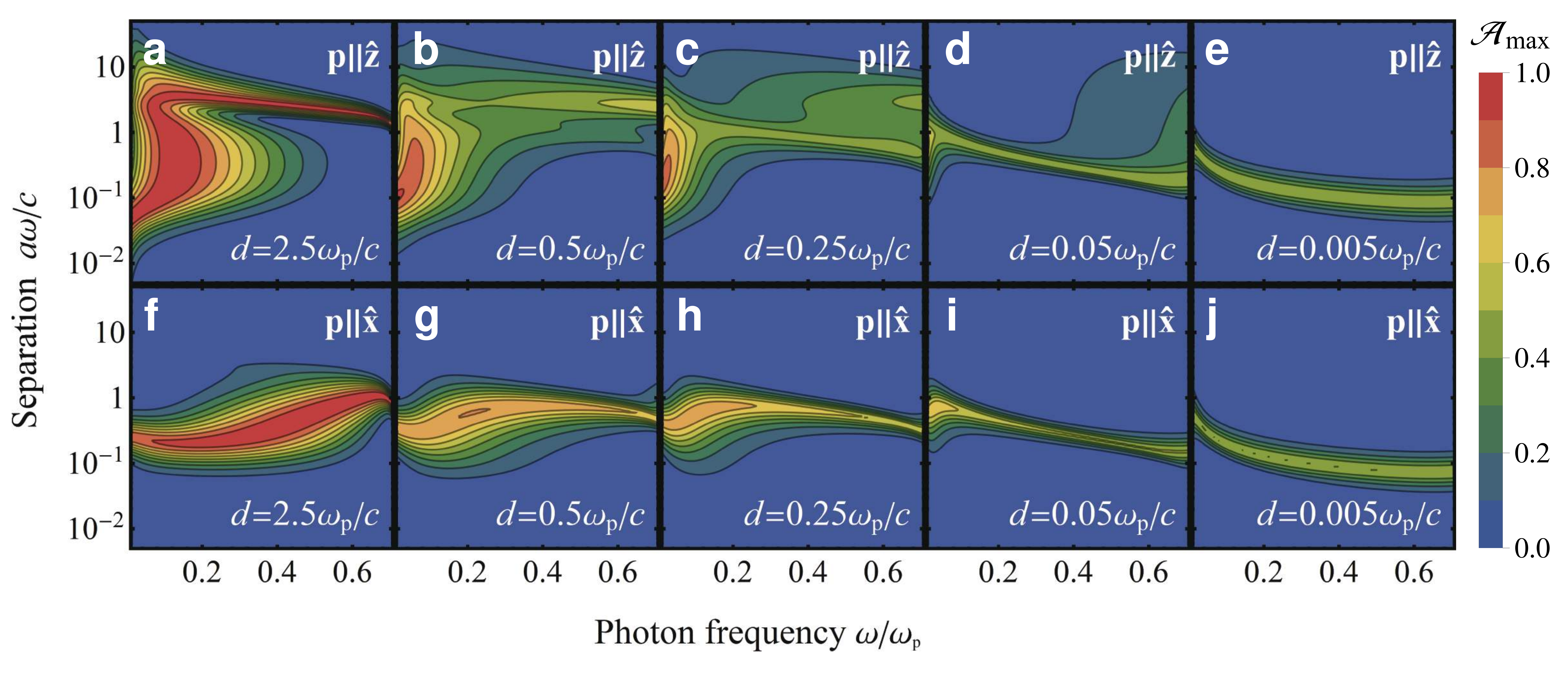}
\caption{{\bf Thickness dependence of optimum coupling to plasmons in Drude metal films.} We show maps similar to those in Fig.\ \ref{Fig2}(a-d) for additional values of the film thickness $d$, as indicated by labels.}
\label{FigS2}
\end{figure*}

\begin{figure*}
\centering
\includegraphics[width=0.9\textwidth]{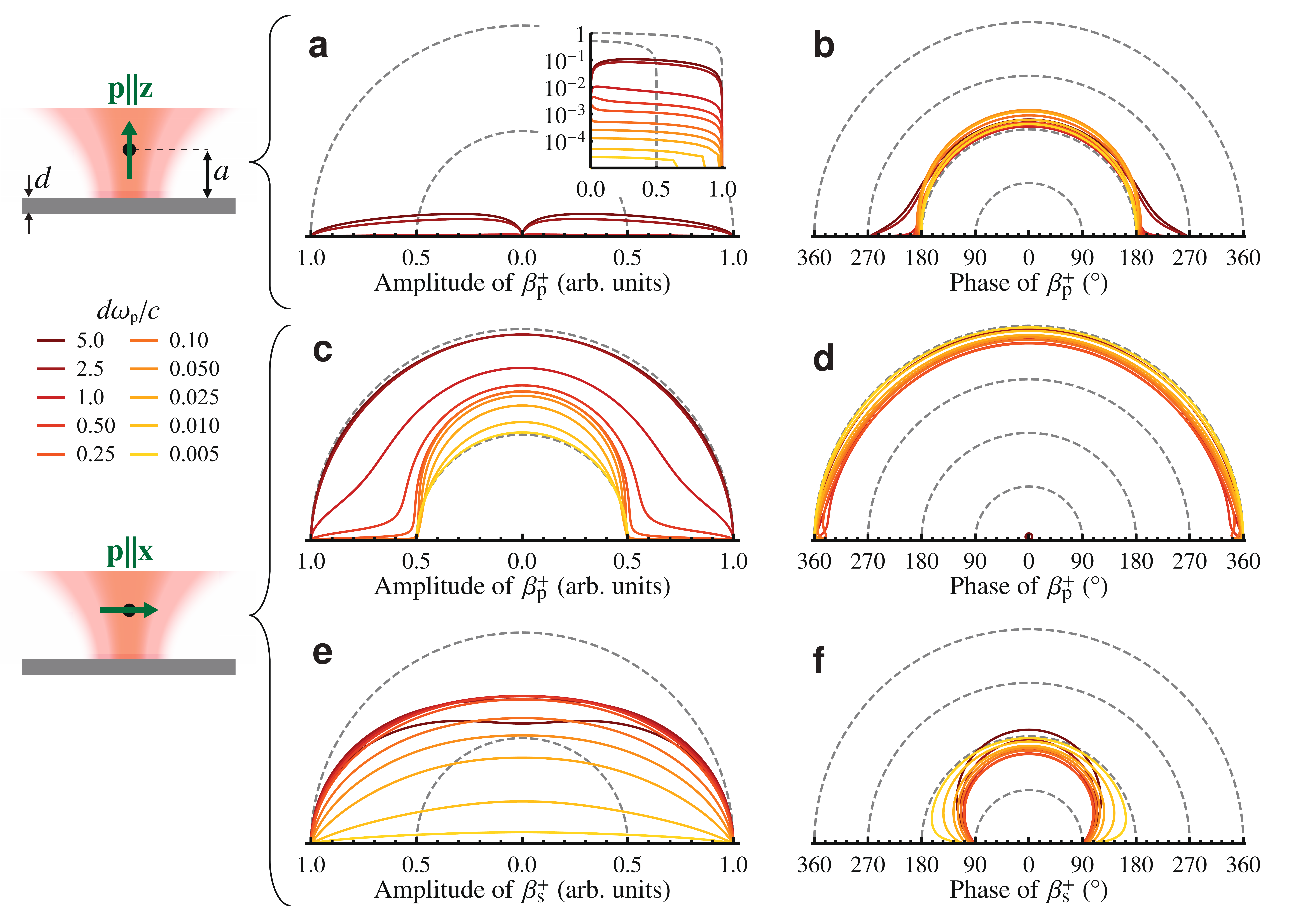}
\caption{{\bf Optimum incident beam profiles for Drude metal films.} Angular distribution of the (a,c,e) amplitude and (b,d,f) phase of the optimized beam parameters $\beta_{\sigma}^+$ under the configuration of Fig.\ \ref{Fig2} for different film thicknesses $d$ (see legend). For each thickness $d$, we choose the corresponding values of $a$ and $\omega$ that maximize the optimized absorption by the film, as prescribed by Fig.\ \ref{Fig2}(e). We consider out-of-plane (a,b) and in-plane (c-f) particle polarizations. The amplitudes of p (c,d) and s (e,f) polarization components are plotted separately for a particle dipole orientation $\pb\parallel\xx$. For $\pb\parallel\zz$, we have $\beta^+_{\rm s}=0$, so this component is not represented. The amplitude for $\pb\parallel\zz$ is also plotted using a linear-log scale in the inset of (a) to improve visibility. As a general trend, when the film thickness decreases, the optimized beam becomes more grazing, while the phase remains rather isotropic.}
\label{FigS3}
\end{figure*}

\begin{figure*}
\centering
\includegraphics[width=0.9\textwidth]{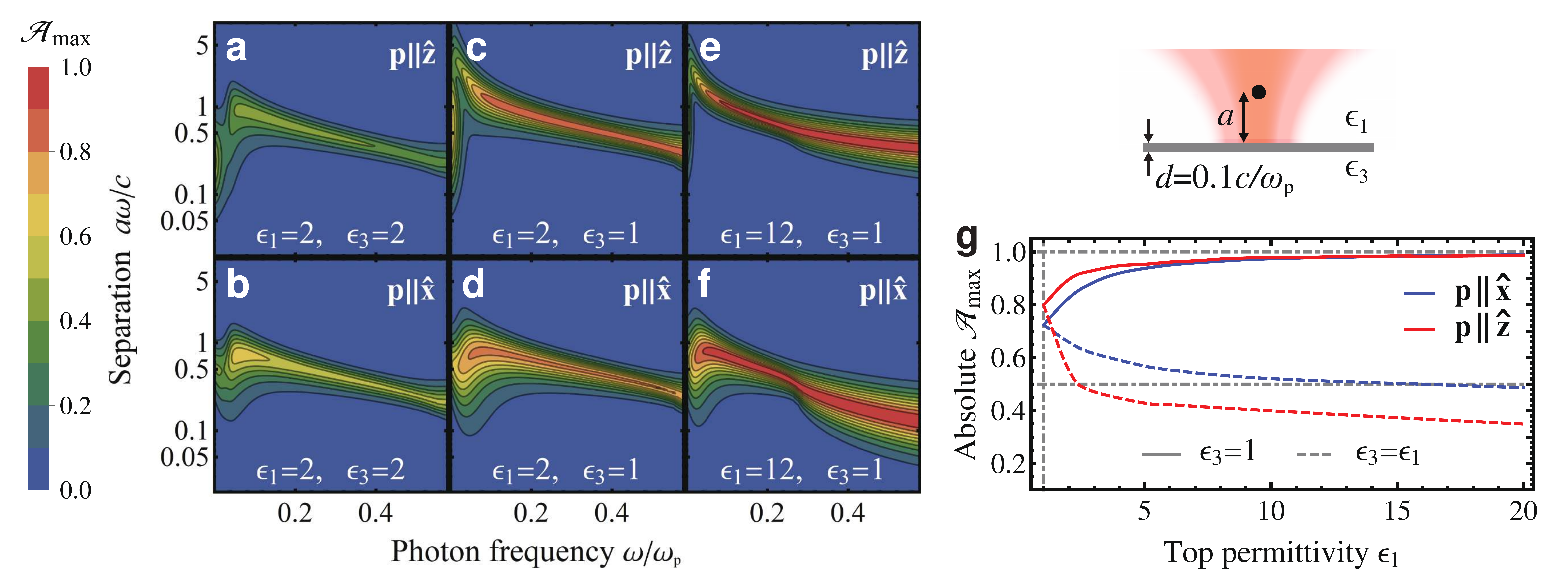}
\caption{{\bf Influence of the dielectric environment in Drude metal films.} (a-h) We represent similar $\omega$-$a$ plots as in Fig.\ \ref{Fig2}(c,d) of the main text for different combinations of the permittivity of the materials in the near (top) and far (bottom) sides of the film (see upper-right sketch). (i) Absolute maximum coupling (assuming a lossless resonant scatterer and optimized with respect to particle-surface distance, light frequency, and incident beam profile) as a function of the host permittivity for symmetric (dashed curves) and asymmetric (solid curves) environments with $\xx$ and $\zz$ particle dipole orientations (see legend). Light is incident from the high-index side in the asymmetric configuration.}
\label{FigS4}
\end{figure*}

\begin{figure*}
\centering
\includegraphics[width=0.75\textwidth]{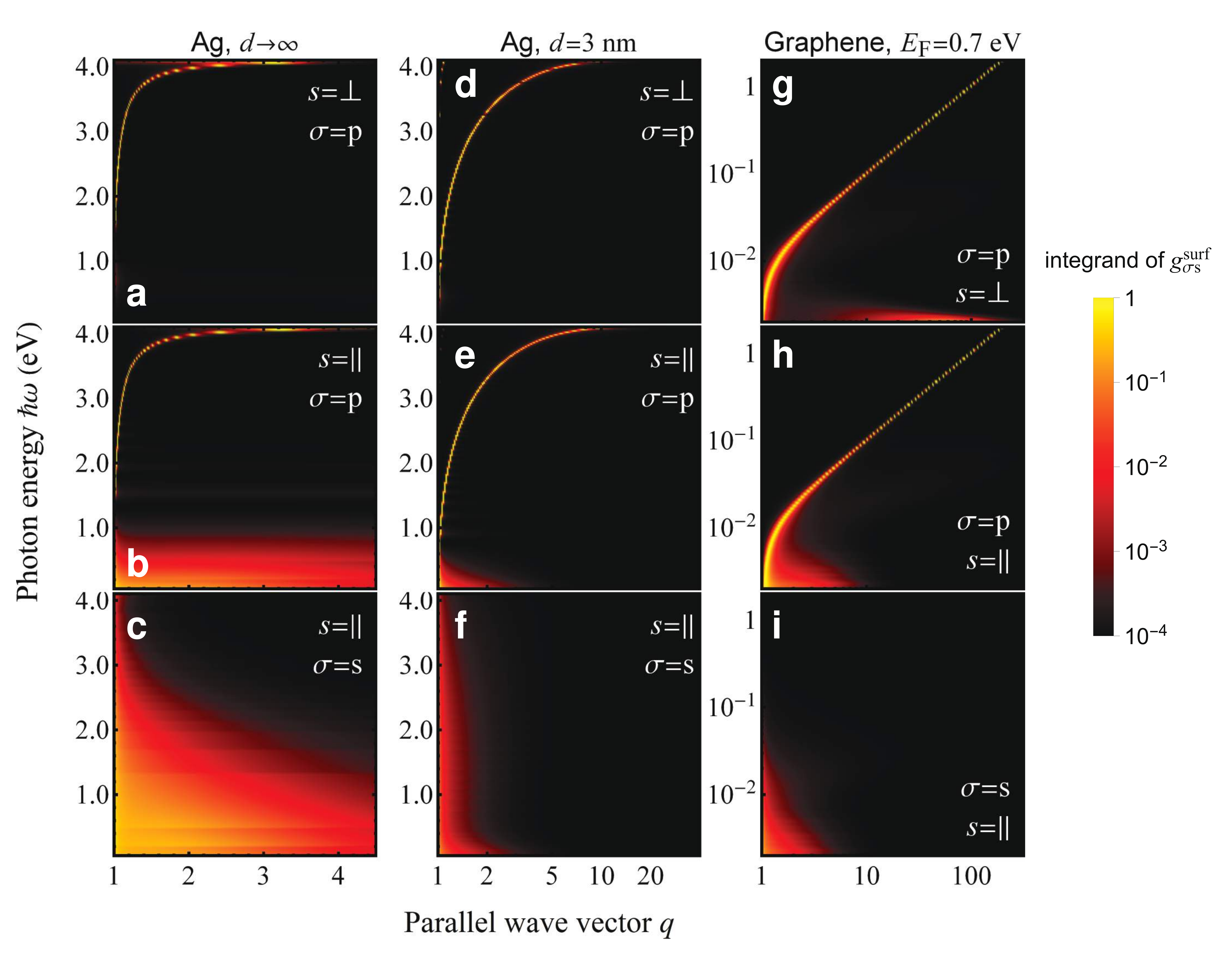}
\caption{{\bf Momentum-resolved contribution to coupling to silver and graphene plasmons .} We plot the integrand of Eqs.\ (\ref{eq:gpar}) and (\ref{eq:gperp}) as a function of $q=\kpar/k'$ outside the light cone for a semi-infinite silver surface (a-c), a thin silver film (d-f), and doped graphene (g-i) in the configurations of $\perp$ (a,d,g) or $\parallel$ (rest) particle dipole orientations. We set the particle-surface distance to the value that maximizes absorption for each frequency [see Figs.\ \ref{Fig2}(e) and \ref{Fig4}(d,e)]. For $\parallel$ orientation, we separate the contributions of p (b,e,,h) and s (c,f,i) light polarization components, which are proportional to $\tilde{r}_{\kpar{\rm p}}$ and $\tilde{r}_{\kpar{\rm s}}$, respectively. The sharp feature in the p polarization plots corresponds to the contribution of plasmons in each surface, which dominates light absorption outside the light cone for most frequencies in thin silver and graphene. In contrast, we find that nonresonant s polarization components play a leading role in the semi-infinite metal for $\parallel$ particle dipole orientation, specially at low frequencies.}
\label{FigS5}
\end{figure*}

\begin{figure*}
\centering
\includegraphics[width=0.9\textwidth]{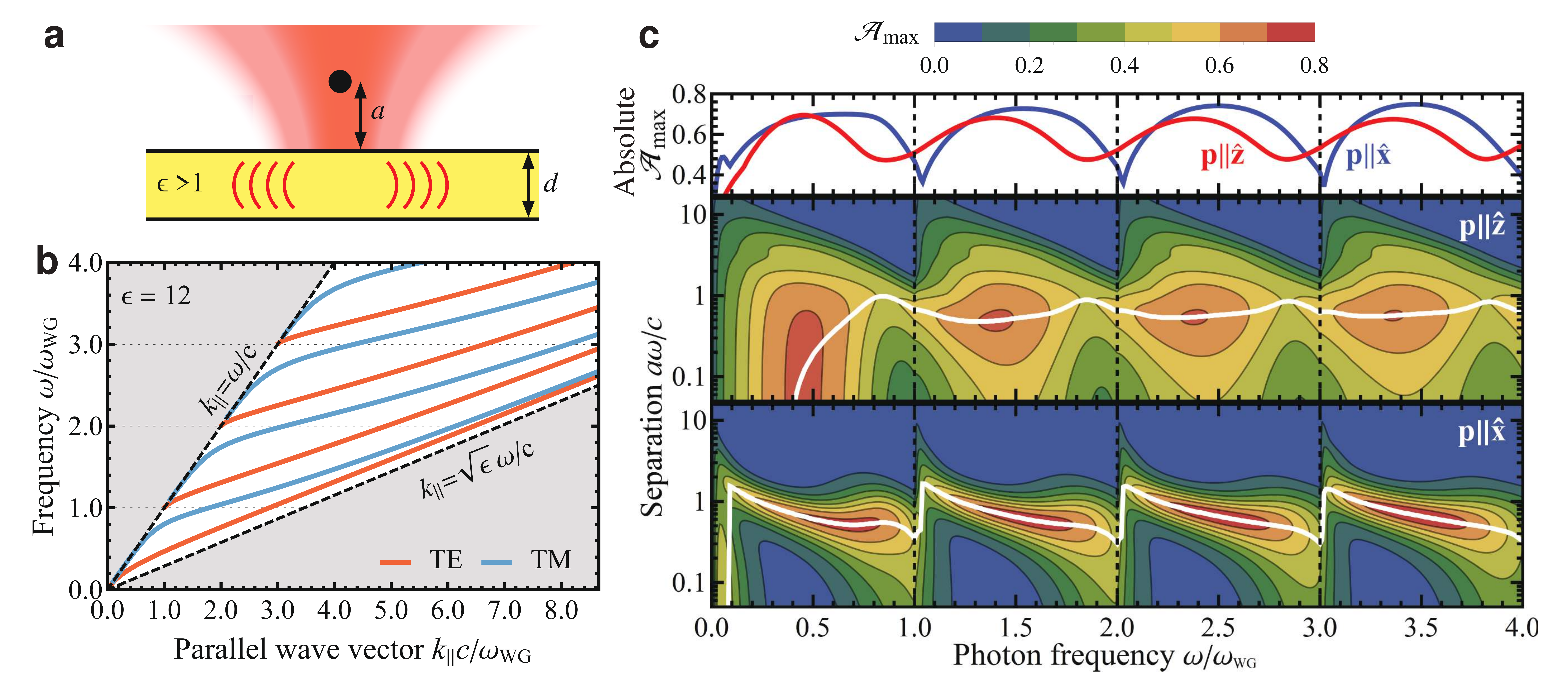}
\caption{{\bf Complete coupling to self-standing dielectric waveguide modes.} Same as Fig.\ \ref{Fig3}, but for self-standing waveguides. Panels (b) and (c) are calculated for $\epsilon=12$. The dispersion relation in panel (b) is substantially different from the one in Fig.\ \ref{Fig3}(b) for a PEC-supported waveguide, and in particular, the lowest-order mode has different polarization. Remarkably, even without the PEC mirror, we find that $>70\%$ of the incident light can be coupled into waveguide modes, regardless of film thickness.}
\label{FigS6}
\end{figure*}

\begin{figure*}
\centering
\includegraphics[width=0.9\textwidth]{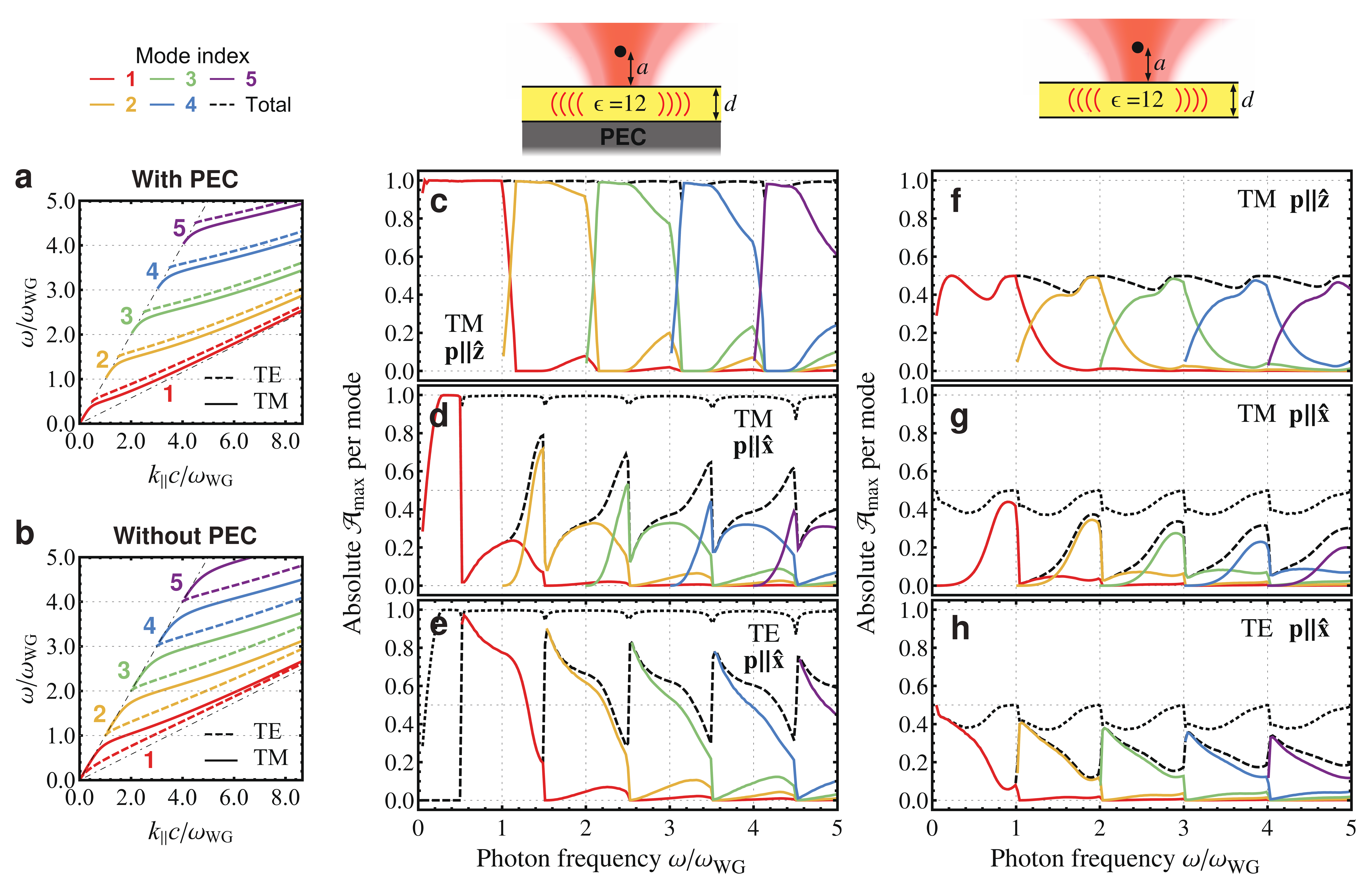}
\caption{{\bf Mode decomposition of the coupling to dielectric waveguides.} (a,b) Dispersion relation of PEC-supported (a) and self-standing (b) $\epsilon=12$ dielectric waveguides of thickness $d$ with the axes normalized using the characteristic frequency $\omega_{\rm WG}=(\pi c/d)/\sqrt{\epsilon-1}$. TM (solid curves) and TE (dashed curves) modes are labeled 1-5. (c-h) Partial contribution of the coupling fraction to different waveguide modes (color-coded solid curves) under conditions of maximum overall coupling (i.e., optimized with respect to incident beam profile, particle polarizability, and particle-surface distance) for $\xx$ and $\zz$ particle dipole orientations (see labels) in the supported (c-d) and self-standing (f-h) film configurations. We show the sum of all contributions within each plot as dashed curves, as well as the sum of all TE and TM modes (dotted curves) in both (d,e) and (g,h).}
\label{FigS7}
\end{figure*}

\begin{figure*}
\centering
\includegraphics[width=0.7\textwidth]{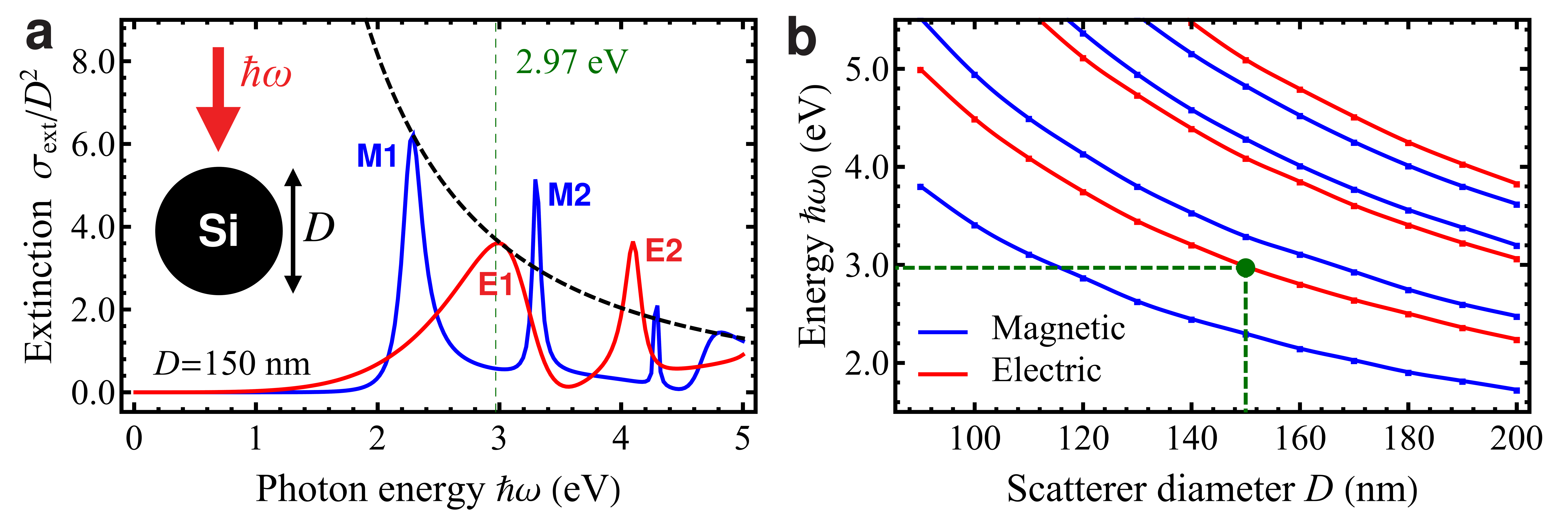}
\caption{{\bf Mie resonances of Si nanospheres.} (a) Magnetic (M) and electric (E) contributions to the scattering cross section of a Si ($\epsilon=12$) sphere of diameter $D=150\,$nm. Dipolar (1) and quadrupolar (2) Mie resonances are identified by labels. (b) Position of the lowest-frequency resonances as a function of Si sphere diameter. Green-dashed lines correspond to the conditions used in Figs.\ \ref{Fig1} and \ref{Fig5} ($D=150\,$nm, $\hbar\omega=2.97\,$eV). We calculate the extinction cross section from Mie theory \cite{M1908} using the expression $\sigma^{\rm ext}(\omega)=(\lambda_0^2/2\pi\epsilon_{\rm h})\sum_{l=1}^{\infty}(2l+1)\,{\rm Im}\{t_l^{\rm E}+t_l^{\rm M}\}$, where \\ \\
$t_l^\nu= \big\{-\eta_\nu j_l(\rho_{\rm s}) [j_l(\rho_{\rm h})+\rho_{\rm h} j'_l(\rho_{\rm h})]+j_l(\rho_{\rm h})[j_l(\rho_{\rm s})+\rho_{\rm s}j'_l(\rho_{\rm s})]\big\}\big{/}\big\{\eta_\nu h_l^{(+)}(\rho_{\rm s})[j_l(\rho_{\rm h})+\rho_{\rm h} j'_l(\rho_{\rm h})]-j_l(\rho_{\rm h})[h_l^{(+)}(\rho_{\rm s})+\rho_{\rm s} h_l^{(+)'}(\rho_{\rm s})]\big\}$ \\ \\
are electric ($\nu=$E) and magnetic ($\nu=$M) scattering matrix elements corresponding to the choice $\eta_{\rm E}=\epsilon/\epsilon_{\rm h}$ and $\eta_{\rm M}=1$, respectively, $\rho_{\rm h}=\pi\sqrt{\epsilon_{\rm h}}D/\lambda_0$, $\rho_{\rm s}=\pi\sqrt{\epsilon}D/\lambda_0$, $\epsilon_h$ is the host permittivity (here set to 1), $h_l^{(+)}$ and $j_l$ are spherical Hankel and Bessel functions, and the primes indicate the derivative with respect to the argument. 
For nonabsorbing spheres, the scattering matrix elements satisfy the analytical property ${\rm Im}\{t_l^\nu\}\le1$, where the equal sign corresponds to resonance conditions, which are always met at some frequency. The extinction cross section at those points is then $\gtrsim(2l+l)\lambda_0^2/2\pi\epsilon_{\rm h}$ (modes other than the resonant one can also produce additional scattering). In particular, the resonant contribution to the extinction cross section by a resonant dipolar mode ($l=1$) is $3\lambda_0^2/2\pi\epsilon_{\rm h}$, which coincides with the maximum cross section for a dipolar scatterer and is plotted in (a) as a dashed curve; in the high-index sphere under consideration, modes of the same polarization (E or M) are well spaced, and consequently, the dashed curve passes close to the peaks of the E1 and M1 resonances.}
\label{FigS8}
\end{figure*}

\begin{figure*}
\centering
\includegraphics[width=0.9\textwidth]{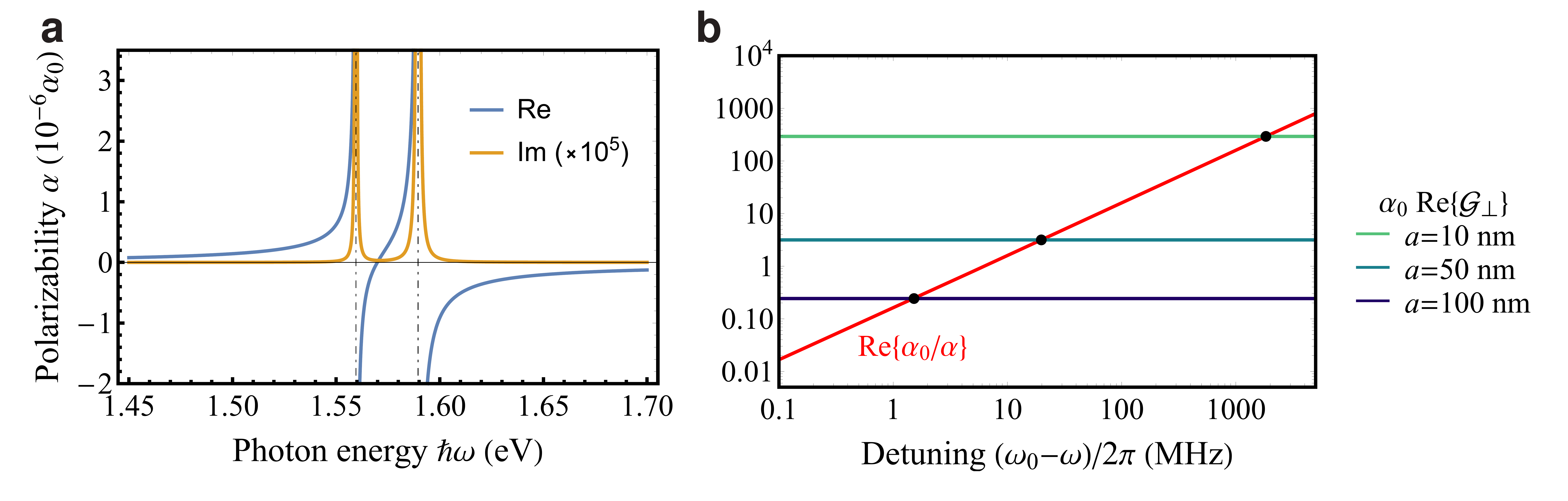}
\caption{{\bf Polarizability of Rb and resonant coupling to Si waveguides.} (a) Real and imaginary parts of the polarizability $\aE$ of a Rubidium atom placed in vacuum and normalized to the reference value $\alpha_0=3/2k^3$. The imaginary part is multiplied by $10^5$ to improve visibility. (b) Comparison of ${\rm Re}\{1/\aE\}$ and ${\rm Re}\{\mathcal{G}_\perp\}$ near the lowest-frequency atomic resonance, with the latter calculated for the scatterer placed at a distance $a$ from the surface of a PEC-supported Si film (see legend). The intersection points (symbols) mark the resonance condition ${\rm Re}\{1/\aE-\mathcal{G}_\perp\}=0$ [Eq.\ (\ref{condition2})]. The film thickness ($d=126$, 187, and 137\,nm for $a=10$, 50, and 100\,nm, respectively) is chosen to fulfil the resonance condition at the 1.56\,eV line of Rb. We model the Rb atomic polarizability using the fitted expression \cite{SWC03} $\aE(\omega)=(3c^3/4)\sum_{j=1,2} (\Gamma_j/\omega_j^3)/(\omega_j-\omega-\ii\Gamma_j/2)$ with parameters $\hbar\omega_1=1.556\,$eV, $\hbar\omega_2=1.589\,$eV, $\Gamma_1=25.1\,$MHz, and $\Gamma_2=12.6\,$MHz.}
\label{FigS9}
\end{figure*}



\end{document}